\documentclass[reprint,twocolumn,preprintnumbers,superscriptaddress,showpacs]{revtex4-1}
\usepackage{amsmath, mathrsfs, amssymb,amsfonts,amsthm,graphicx, epsf, dcolumn, yfonts}
\usepackage[hyperfootnotes=true]{hyperref}
\usepackage{subfigure}
\usepackage{color}
\usepackage{slashed}
\usepackage{setspace}
\usepackage{cancel}
\usepackage{wasysym}
\usepackage{hyperref}
\pdfoutput=1
\usepackage{float}

\newcommand{\ba}{\begin{eqnarray}}
\newcommand{\ea}{\end{eqnarray}}

\newcommand{\la}[1]{\label{#1}}
\newcommand{\eq}[1]{(\ref{#1})}

\newcommand{\be}{\begin{equation}}
\newcommand{\ee}{\end{equation}}
\newcommand{\bea}{\begin{eqnarray}}
\newcommand{\eea}{\end{eqnarray}}

\def\pp{\partial}

\def\a{\alpha}

\def\s{\sigma}

\def\lab{\label}

 \def\cH{{\cal H}}

\newcommand{\vev}[1]{{\left< {#1} \right>}}

\newcommand{\prt}[1]{{\left( {#1} \right)}}

\renewcommand*{\thefootnote}{\fnsymbol{footnote}}

\begin{document}

\title{Neural Network flows of  low $q$-state Potts and clock Models} 

\author{Dimitrios Giataganas}
\email{dgiataganas@phys.uoa.gr}
\affiliation{Department of Physics, University of Athens, \\ 
Zographou 157 84, Greece}

\author{Ching-Yu Huang}
\email{ayajor827@gmail.com}
\affiliation{Department of Applied Physics, \\
Tunghai University, Taichung 40704, Taiwan}

\author{Feng-Li Lin}
\email{fengli.lin@gmail.com}
\affiliation{Department of Physics, \\
National Taiwan Normal University, Taipei, 11677, Taiwan}
\affiliation{Center of Astronomy and Gravitation, \\
National Taiwan Normal University, Taipei 11677, Taiwan}

\begin{abstract}
It is known that a trained Restricted Boltzmann Machine (RBM) on the binary Monte Carlo Ising spin configurations, generates a series of iterative reconstructed spin configurations which spontaneously flow and stabilize to the critical point of physical system. Here we construct a variety of Neural Network (NN) flows using the RBM and  (variational) autoencoders, to study the $q$-state Potts and clock models on the square lattice for $q=2,3,4$. The NN are trained on Monte Carlo spin configurations at various temperatures. We find that the trained NN flow does develop a stable point that coincides with critical point of the $q$-state spin models. The behavior of the NN flow is nontrivial and generative, since the training is unsupervised and without any prior knowledge about the critical point and the Hamiltonian of the underlying spin model. Moreover, we find that the convergence of the flow is independent of the types of NNs and spin models,  hinting a universal behavior.  Our results strengthen the potential applicability of the notion of the NN flow in studying various states of matter and offer additional evidence on the connection with the Renormalization Group flow.
\end{abstract}

\renewcommand*{\thefootnote}{\arabic{footnote}}
\setcounter{footnote}{0}

\maketitle

\section{Introduction}
\lab{sec:intro}

Machine learning (ML) methods outperform humans in specific tasks and have been applied successfully in a wide range of modern physics  \cite{manybody1,troyer1,Hush:2017,Cai:2018,Torlai:2018,exotic1,fermionsign1}.  Of particular interest are  applications of ML on identifying and classifying different phases of matter including the topological ones  \cite{spinlic2,manybodyloc,looptopography,huber1,Nieuwenburg:2018,Suchsland:2018,glassy1,giataganas1,Iso:2018yqu,Koch:2019fxy,Ohtsuki:2019qnk,Alexandrou:2019hgt,review1}. The identification of the phase transitions of spin systems, is in practice an accessible task where the spin states can be mapped to neural network states and in a sense one retains  physical information during the training of the neural network.  There are several methods that the neural network (NN) have been used to identify the phase transitions of the spin models including supervised and unsupervised training. A particularly interesting one is the so called Restricted Boltzmann  Machine \cite{Hinton:2012} (RBM) flow introduced and studied in \cite{Iso:2018yqu,giataganas1}. The idea is to train an RBM on the Ising spin binary states produced by Monte Carlo simulations at various temperatures and external magnetic fields. Once the training is done and the parameters of the RBM are fixed, one defines an RBM flow as the sequence of binary image reconstructions by the RBM, with the initial input been a Monte Carlo spin configuration at a certain temperature and magnetic field. It has been found that the RBM flow indeed generates spin states at various temperatures in a certain direction, towards a stable point. The stable point of the RBM flow turned out to match with the one that maximizes certain thermodynamic quantities like the specific heat in the Ising model \cite{giataganas1}. In the absence of the external magnetic field the maximization and divergences of the thermodynamic quantities happens at the critical fixed point of the system where the phase transition happens. Therefore the RBM flow does identify spontaneously the phase transition of the Ising model, and resembles partly a process of the Renormalization Group (RG) flow on physical system \cite{Iso:2018yqu}.

The novelty of the RBM flow is that the identification of the phase transition under the flow happens in a completely spontaneous way. The RBM has no prior knowledge about the physical system such as the Hamiltonian and its phase structures because we do not label the input microstates when training the RBM. Moreover, the spin configurations of the training set is not biased in any way towards the critical ones. Thus, it is quite remarkable that the RBM can flow all the microstates to the ones at the critical point. This unsupervised feature is in contrast to other supervised learning of the phase transitions \cite{spinlic2,manybodyloc,looptopography,huber1,Nieuwenburg:2018,Suchsland:2018,glassy1,giataganas1,Iso:2018yqu,Koch:2019fxy,Ohtsuki:2019qnk,Alexandrou:2019hgt,review1}. A further impressive application of the RBM flow is that its fixed-point microstates, has been used to compute successfully the critical exponents of the spin models \cite{giataganas1} and the scaling dimensions of operators \cite{Koch:2019fxy}. In addition, the spontaneous nature of RBM flow, can be seen as a strong hint on the existence of the fundamental relation between the way that the learning occurs in neural networks and the RG flow in statistical physics \cite{Beny:2013,mehta1,Paul:2014,Aoki:2016,Tegmark,foreman,ringel1,Hashimoto:201802,Chung:2020jyo,koch2020unsupervised}. The connection between RG and ML relies on K. G. Wilson's idea behind the RG flow \cite{wilson1}, where by identifying relevant degrees of freedom and integrating out the irrelevant a low energy effective theory with universal features is obtained. This kind of mechanism can be thought as exploited by artificial systems, where the neural layers perform the local operations.
In this sense, we expect that the NN other than RBM should be also capable of generating the so-called NN flow to identify the phase transitions spontaneously.

In this work we extend the studies of the RBM flow developed in \cite{giataganas1,Iso:2018yqu} by two new aspects. The first is going beyond the binary lattices to work with the q-state Potts models or their variations such as clock models or called planar Potts models. We are studying physical systems with richer structure and information compared to Ising model. To utilize the binary feature of the RBM, we need to encode the value of a single q-state spin into several binary nodes of RBM's visible layer. As a result, the neural network representation of the physical data becomes more nontrivial than the cases of binary lattices. In some sense, the space of physical microstates is just the sub-space of the full neural network because only part of the input-layer binary vectors of RBM are used as the physical inputs. This is a main difference compared to the Ising model and previous works on the RBM flow. Another new aspect of the extension to \cite{giataganas1,Iso:2018yqu} is to realize the similar flow by using the NN other than RBM. In general we call the machinery the NN flow. In this paper we consider the NN with the structure of either autoencoder (AE) or the variational autoencoder (VAE). In contrast to RBM, one advantage is that these NNs are not restricted to be binary, and should be easier to implement  the training for the q-state lattices. Our results reveal that for each NN considered, the flow identifies spontaneously the critical point of the Potts model in its various versions, based exclusively on training by the states of the system.

As a side note we mention that such multistate models have various applications even beyond physics. The tumor growth, the properties of cell sorting,  the patterns of collective segregation and the rheological properties of foams, are just few of the popular cross field applications of Potts model \cite{graner,schelling,Szab2013CellularPM,foam}. Our methods could be potentially useful to identify the critical points in such variations of multistate models, wherever they play an important role in the study.   We also remark that different types of applications of the NN on the Potts model and its phase transitions have been considered recently in \cite{Shiina_2020,Tan:2020cha,Li:2017xaz}.

Our paper is organized as following. In section \ref{sec:review}, we present a brief introduction of the $q$-state Potts and clock model, and the associated phase diagrams for $q=2,3,4$ obtained form our Monte Carlo spin configurations. In section \ref{sec:method}, we sketch the structure of the NN flow, and elaborate the details of each components and their training procedures for both RBM and (V)AE flows. In the section \ref{sec:results} we present the results of the RBM, AE and VAE flows in identifying the critical points shown in section \ref{sec:review}, and also discuss their implications, and the related issues on the effects of finite-size and magnetic field. We then conclude our papers with summary and discussions in the section \ref{sec:con}. In Appendix we show the typical views of the spin configurations of the input, latent and output layers of (V)AE flow, and the typical training loss and accuracy.

\begin{figure*}[htb!]
  \centering
    \subfigure[ ]{
    \label{phase-q2-clock} 
    \includegraphics[width=0.31\textwidth]{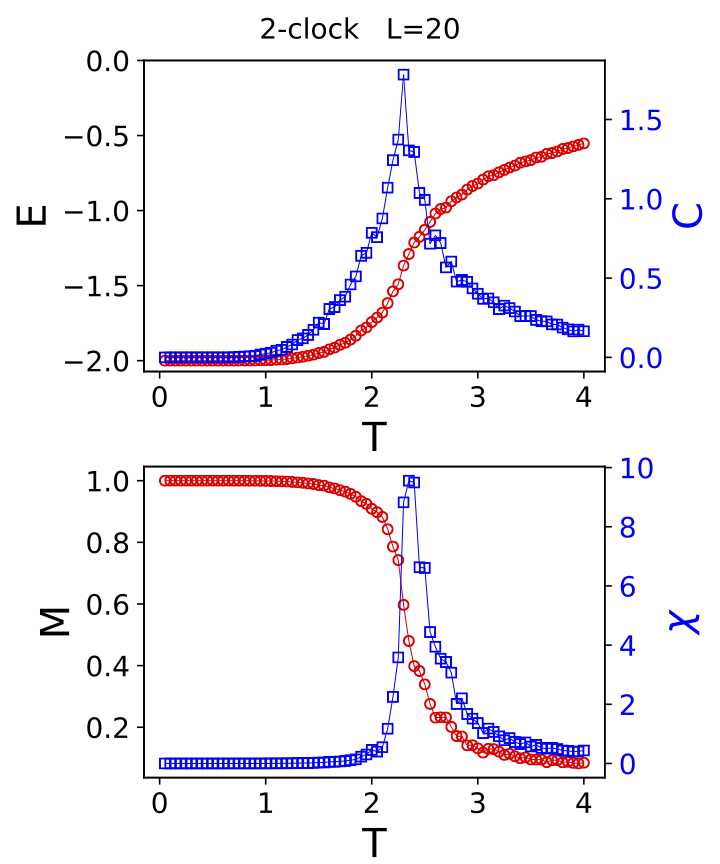}}
    \subfigure[ ]  {
    \label{phase-q3-Potts} 
    \includegraphics[width=0.31\textwidth]{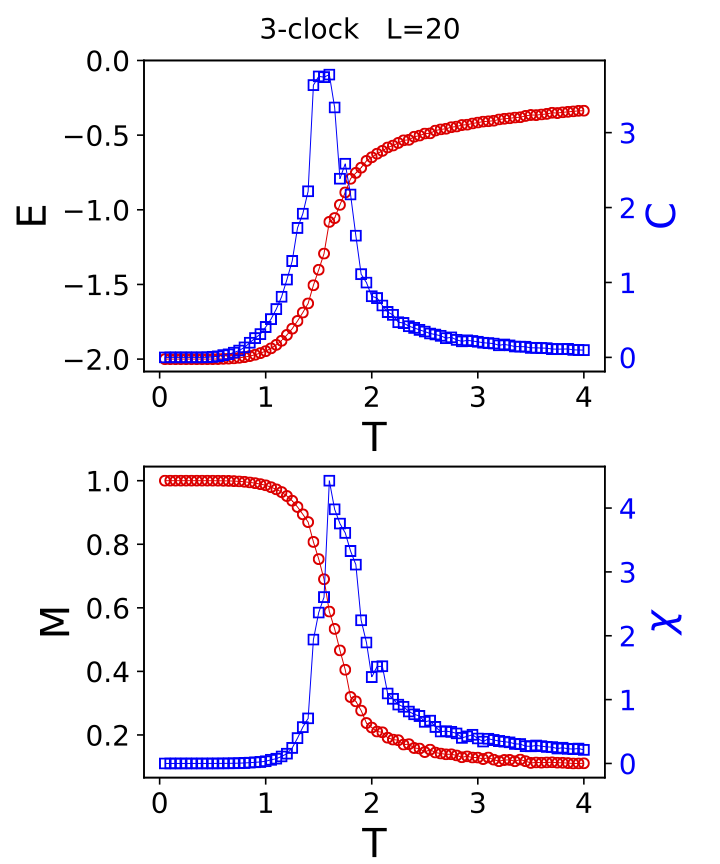}}
    \subfigure[ ]  {
    \label{phase-q4-Potts} 
    \includegraphics[width=0.31\textwidth]{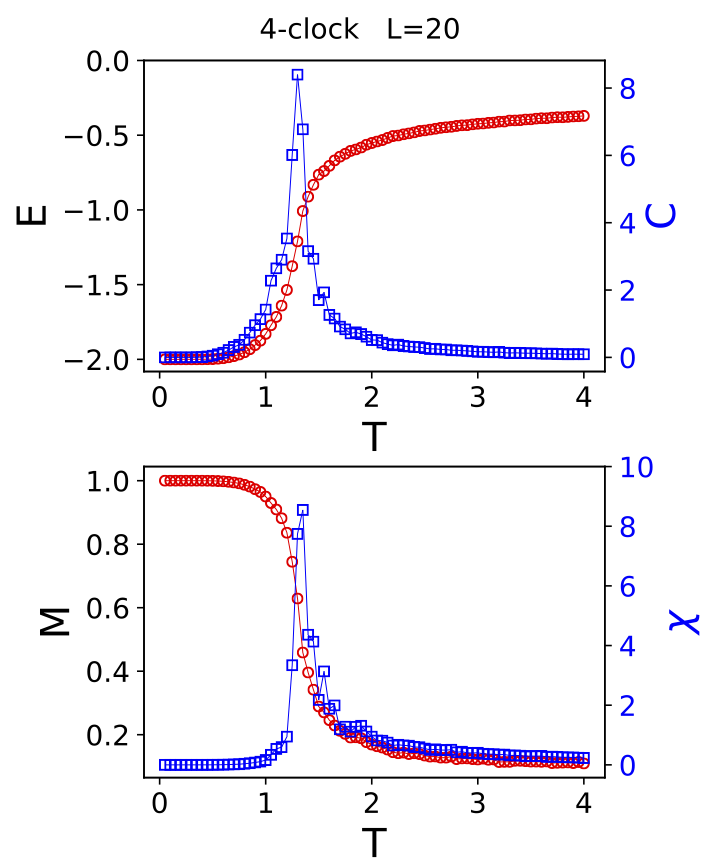}}
  \caption{ (a)  $q=2$ (b)  $q=3$ (c)  $q=4$  clock models of 20 by 20 sites. {\bf (i) [Top-red]} $E$ (Energy) vs $T$ (Temperature),
  {\bf (ii) [Top-blue]} $C$ (Specific heat) vs $T$, (iii) {\bf [Bottom-red]} $M$ (Magnetization vs) $T$, and {\bf (iv) [Bottom-blue]} $\chi$ (Magnetic susceptibility) vs $T$. These graphs indicate a critical temperature around (a) $T_c\simeq 2.3$ for $q=2$, (b) $T_c\simeq 1.50$ for $q=3$, and (c) $T_c\simeq 1.20$ for $q=4$. Note these MC critical temperatures are slightly different from the theoretical ones listed in Table \ref{table: critical_point}.
    \label{phase diagrams}  }
\end{figure*}

\section{The $q$-state spin models}\lab{sec:review}

The Hamiltonian of the $q$-state spin model on the square lattice is  \cite{pottsmodel}
\be\la{hamilton}
\cH=-J \sum_\vev{ij} f\prt{\theta_i , \theta_j}~,
\ee
where $\vev{ij}$ denotes the sum over all the nearest neighbor pairs of the sites $i$ and $j$. $J>0$ is the nearest-neighbor coupling and the value of the Potts variable $\theta$ on each site is given by $2 \pi \s/q$ with site spin variable $\s=0, \ldots, q-1$. The partition function at finite temperature $T$ is defined
as
\be
Z_N=\sum_{\{\s\}} \exp\prt{ K \sum_\prt{i,j}f\prt{\theta_i,\theta_j}}~,
\ee
where the sum $\{\s\}$ is among all the possible spin configurations and $K:=J/T$ is the normalized coupling with respect to $T$, while the Boltzmann constant is set equal to unit, $k_B=1$.

The standard choice of the function $f$ in Hamiltonian \eq{hamilton} is the Kronecker delta $f\prt{\theta_i , \theta_j}:=\delta\prt{\theta_i , \theta_j}$. This choice is canonical and is called the standard Potts model, to which we will refer here simply as Potts model. The critical points of Potts model are known for all $q$. Another popular choice of the interaction is $f\prt{\theta_i , \theta_j}:=-\cos\prt{\theta_i - \theta_j}$. This model is called the clock model or planar Potts model. Obviously, the Potts model and clock model are equivalent for $q=2$ up to a rescaling of the coupling $K$ by a factor of two. These 2-state models can be further shown to equivalent to zero external field Ising model by rewriting their coupling function as $f:=\prt{1+\s_i \s_j}/2$, and by re-scaling of $K$, so that the \eq{hamilton} gives the usual Ising Hamiltonian. Additionally, for $q=3$ the Potts and clock models can be shown to be equivalent by the re-scaling $K \rightarrow 3 K/2$. For $q>4$ there is no apparent relations between these models.

The Potts model can only be solvable for the whole coupling space in the cases of an Ising spin chain or an Ising lattice. At certain corners of the coupling space, there is additional solvability, for example $K=-\infty$ is also solvable for $q=3$ and equivalent to the three-colouring problem. Nevertheless the critical point of the q-state Potts model in two-dimensional lattice is known to be $K_c=\log\prt{1+\sqrt{q}}$ for all $q$, which means that by setting the coupling $J=1$ the  critical temperature reads
\be
 T_c=1/\log\prt{1+\sqrt{q}}\;.
\ee
This is a statement based on the vertex model equivalence of the Potts model, where the zeros of the partition function occur only when the above relation is satisfied. The rigorous proof is valid only for $q\ge 4$, but expected to work for lower values since it reproduces well the $q=2$ prediction of the Onsager's solution. The critical points of Potts models are of second order for $q\le 4$, and are of first order for $q>4$. On the other hand, the critical points of the q-state clock models are of second order only for $q\le 4$,  whose critical temperatures \cite{ORTIZ2012780} are listed in Table \ref{table: critical_point} along with the ones of Potts models for comparisons. For $q>4$, the critical points of the clock models are the Berezinskii–Kosterlitz–Thouless (BKT) transition \cite{Lapilli_2006,Li_2020}, which was first proposed in classical $XY$ model with a continuum $U(1)$ symmetry \cite{Kosterlitz:1973xp}.
\renewcommand{\arraystretch}{1.1}
\begin{table}[htb]
	\centering
\begin{tabular}{ |c| c| c|  }
\hline
 Types & Potts & Clock \\ \hline
 \hline
q=2  &   $T_c \approx 1.135$ &  $T_c= \frac{2}{\ln{(1+\sqrt{2})}} \approx 2.269$ \\ \hline
q=3  &  $T_c \approx 0.995 $ & $T_c= \frac{3}{ 2\ln{(1+\sqrt{3})}} \approx 1.492$  \\ \hline
q=4  &  $T_c \approx 0.910 $ & $T_c= \frac{1}{\ln{(1+\sqrt{2})}} \approx 1.135$  \\  \hline
\end{tabular}
\caption{The critical temperatures ($T_c$) for the $q$-state Potts and clock models with $q=2,3,4$.  }
\label{table: critical_point}
\end{table}

\subsection{Monte-Carlo on Potts and clock models} \label{sec:MC simulations}

We implement the Monte-Carlo (MC) method to simulate the spin configurations of the Potts and clock models for a range of temperatures, and use them as the inputs to train the unit for NN flow. Specifically, we use the Metropolis Monte Carlo  simulation to generate $q=2,3,4$ spin configurations on a square lattice for the temperatures ranging from $T=0$ to $T=4$ with increment by $\Delta T=0.05$. We generate $1000$ (or $2000$ for some cases) MC spin configurations for each temperature. The size of the square lattice is specified for each training below, and for certain cases we use different lattice sizes for checking the dependence of our results on the lattice size. A representative lattice size is $20$ by $20$, where the finite size effect is relatively small and the computational time required to create the data set is not long. Moreover, we have use the toroidal boundary conditions for our simulations but we do not expect our results to depend on this detail. The coupling is fixed in the rest of the paper to $J = 1$ without loss of generality. The majority of our MC simulated spin configurations are used as the training data, and the remaining for validation to soften the over-fitting when training the unit of NN flow. 
 
\begin{figure*}[t!]
\includegraphics[angle=0,width=0.9\textwidth]{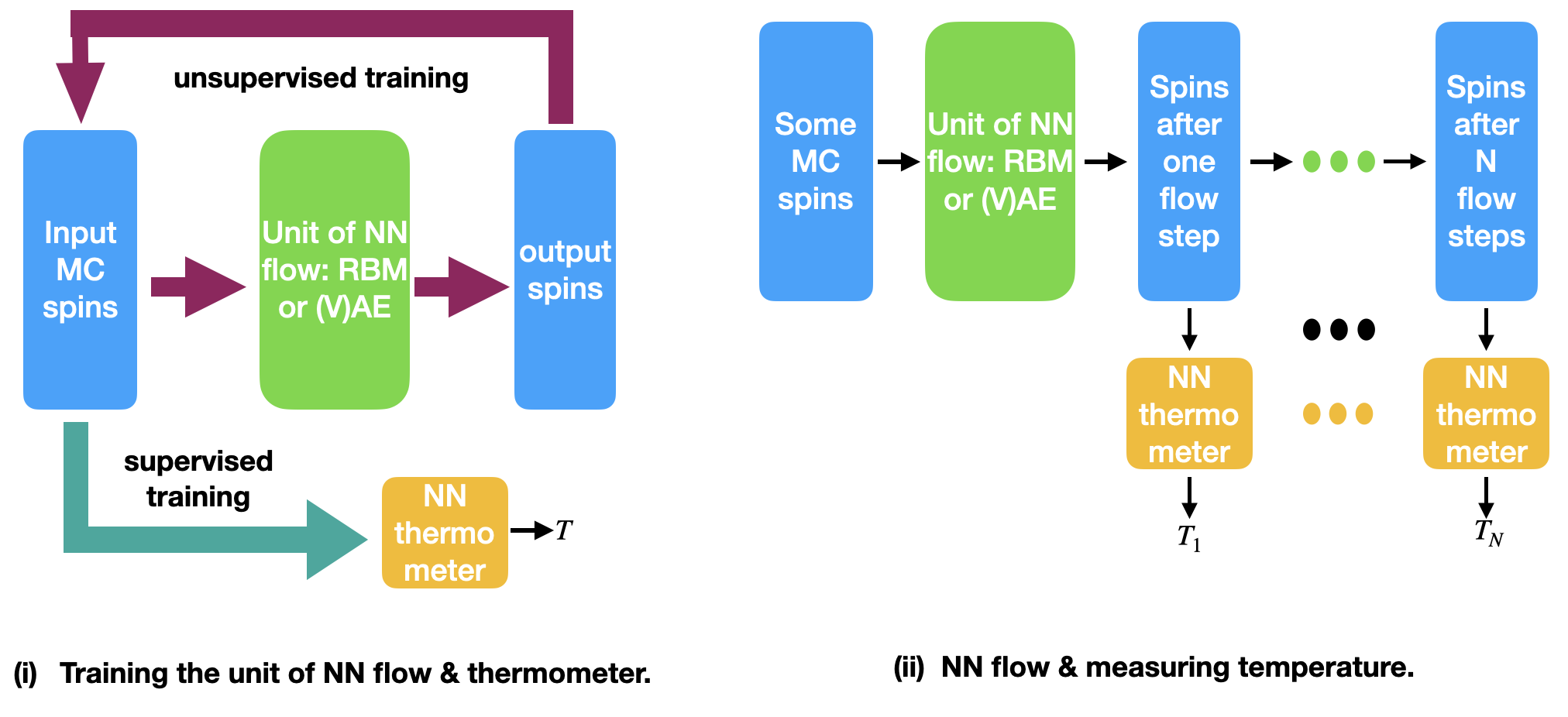}
\caption{\small Summary of NN flow and its training. The step (i) is to train the unit of NN flow and the NN thermometer. The step (ii) is to run the NN flow and then measure the temperatures of the flowed spin configurations. In this way, we can trace the NN flow by the changes of the temperature distribution.}
\label{fig:Summary_of_NN_flow}
\end{figure*}

To characterize the phase transition by thermodynamic observables, we evaluate the energy, specific heat, magnetization and magnetic susceptibility with respect to a set of sampled MC simulations. The results for $q=2,3,4$ clock models are presented in Figure \ref{phase-q2-clock}, \ref{phase-q3-Potts} and \ref{phase-q4-Potts}, respectively. The critical temperatures extracted from the MC simulations are slightly different from the theoretical ones listed in Table \ref{table: critical_point} due to finite size effects.

\section{ Neural Network Flow} \label{sec:method}

In this section we introduce the setup of the various NN we consider for the training and discuss the idea of the NN flow which plays major role in our results. We start by the RBM and then present the AE and VAE neural networks.

The overall NN flow and its key training procedure are outlined in Fig. \ref{fig:Summary_of_NN_flow}. The first step is to perform the unsupervised training of NN-flow unit, and the supervised training of the NN thermometer, both with about $1000$ (or $2000$ for some cases) MC simulated spin configurations. After the training, the second step is to use the NN-flow unit to flow some given MC spin configurations, and then trace the flowed configurations by measuring their temperatures with the trained NN thermometer. The goal is to see if the NN flow can identify the critical point spontaneously. Below we will describe the details for each component and step of Fig. \ref{fig:Summary_of_NN_flow}.

\subsection{Restricted Boltzmann Machines and the flow of reconstructions}

The unit of the NN flow can be either Restricted Boltzmann machines (RBM) or (variational) autoencoder ((V)AE). We first consider the unsupervised learning with the shallow RBM of one layer. The choice is made due to the simplicity of the network that can be used as building block for larger ones and taking into account that such shallow networks perform well on discrete physical models \cite{Morningstar:2017}.

\subsubsection{Brief introduction to RBM}
The Gibbs-Boltzmann probability density for the aforementioned binary and shallow RBM is given be
\be \label{boltzmann1}
p(v_i, h_a) = \frac{e^{-E(v_i, h_a)}}{\cal Z}~,
\ee
where $E$ is the energy function associated to the network
$E(v_i, h_a) = -\sum_{i,a} v_i W_{ia} h_a - \sum_i b^{(v)}_i v_i - \sum_a b^{(h)}_a h_a~,$
$W_{ia}$ is the weight matrix coupling the hidden and visible layers and $b,c$ are the parameters associating the significance of each node in the training, named also as biases.
The $\cal Z$ is the partition function ${\cal Z}=\sum_{\{v_i,h_a\}} e^{-E(v_i, h_a)}$, computed by summing all the possible combinations of the visible and hidden vector states, encoding the exponential complexity of the neural network. In our training method, we will avoid this complexity by approximating this computation on a considerably smaller sampling set. We use a Markov chain sampling to alternate between samples  drawn from the conditional probabilities of each layer which depend on the conditional expectations of the previously sampled layer. This consists of the contrastive divergence (CD) method, and despite that is such brute force approximation which cuts off significant size of the sample information, the training efficiency has been proven very successful \cite{Hinton:2002,Tieleman:2008}.

The RBM parameters $\theta:=\prt{W,b,c}$ are trained by minimizing the distance between the probability distributions of the input Potts state data $q(v_i)$ and the output data $p(v_i)$. The input consists of all possible spin states generated by a Monte Carlo simulation of the Potts model in various temperatures placed randomly in a single training data set. As a measure of the distance between the $q$ and $p$ distributions we choose the relative entropy $\mathrm{KL}(q||p)$, i.e., the Kullback-Leibler (KL) divergence,  and we successively move towards the minimum by renewing the weight matrix and biases, as
$\theta_{i} \to \theta_{i} - \epsilon \frac{\partial\,\mathrm{KL}(q||p)}{\partial \theta_{i}}$. The step of the  gradient descent method $\epsilon$ is set $10^{-3}$ based on experience to optimize the computational time required while achieving to reach to the minimum. The derivatives we minimize are obtained from \eq{boltzmann1}
\bea
\frac{\pp~ \mathrm{KL}(q| |p)}{ \pp W_{ia}}
&=& \langle v_i h_a \rangle_{p(h_a|v_i)} - \langle v_i h_a \rangle_{p(v_i, h_a)}~,\\
\frac{\pp~ \mathrm{KL}(q| |p)}{\pp b_i^{(v)}}
&=& \langle v_i \rangle_{q(v_i)} - \langle v_i \rangle_{p(v_i, h_a)}~, \\
\frac{\pp~ \mathrm{KL}( q| |p)}{\pp b_a^{(h)}}
&=& \langle h_a \rangle_{p (h_a|v_i)} - \langle h_a \rangle_{p(v_i, h_a)}~,
\eea
where $\langle x \rangle_p$ denotes the expectation value of $x$ with respect to the probability density $p$, while $p(h_a | v_i)$ is the probability density of $h_a$ for given $v_i$ with the probability density $q(v_i)$. The computation of the gradients requires an exponential complexity and therefore we approximate the evaluation by running the Markov chain with the probability distribution $p(v_i, h_a)$ in  a step, using as starting point the input data $q(v_i)$. The learning epoch is chosen to be $10^4$.

\begin{figure}[t!]
\includegraphics[angle=0,width=0.3\textwidth]{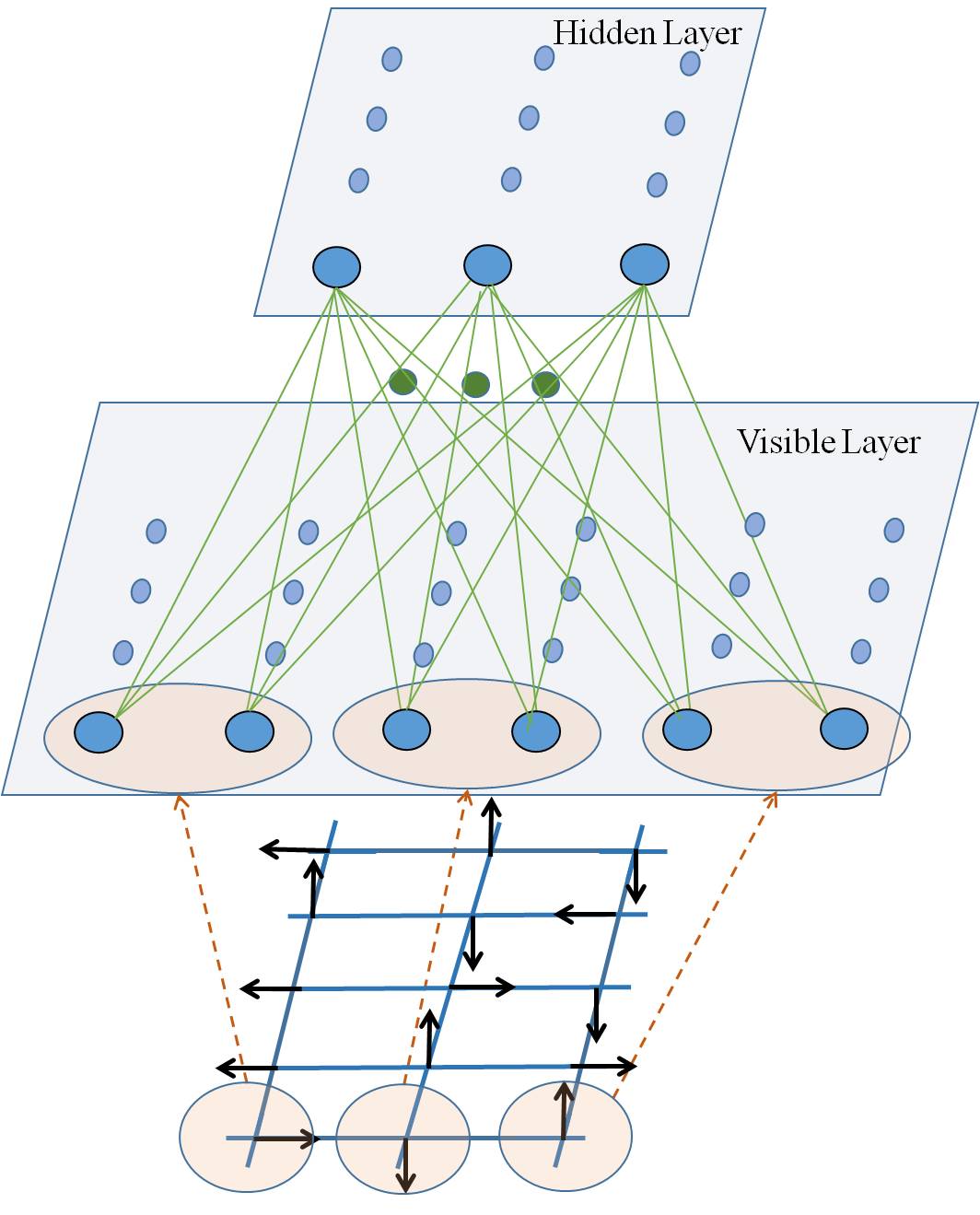}
\caption{\small We map each node of the $q$-state Potts or clock models to $n_{q}$ nodes of visible layer of the RBM, where $n_q$ satisfies $2^{n_q}\ge q > 2^{n_q-1}$. If $2^{n_q}>q$, then the space of the spin configurations is only the sub-space of the space spanned by the visible layer. In this figure we consider the $q=4$ case so that $n_q=2$ so that each q-state spin is encoded into two binary nodes of the visible layer of the RBM.
\label{fig::free}}
\end{figure}

There is an ambiguity on how we represent the Potts state in the Boltzmann machine (and the other NN we use), related to the fact that the Boltzmann machine is a binary model and the Potts in its general case not. The only case that the information of states can be mapped one to one for each node between Potts and RBM is the Ising $q=2$ model. For higher $q$ we decide to encode the information by assigning to each node of the q-state spin model to $n_{q}$ nodes of visible layer of the RBM, where $n_q$ satisfies $2^{n_q}\ge q > 2^{n_q-1}$. An illustration of the way we map the spin model to the RBM is presented in Figure \ref{fig::free}.

\subsubsection{The RBM flow} \la{sub:rbmflow}

Once the training has been performed by the minimization of the KL divergence and the NN parameters have been fixed, we like to find the way to question what is the pattern that the machine has learn. The way we ask this question is to feed a new Potts microstate  $q(v_i)$  generated by the Monte Carlo simulation to our trained neural network and to ask it to reconstruct the image. The output $p(v_i)$ is a new configuration. We may repeat the process iteratively by using as a new input the $p(v_i)$ so we may define a flow of probability distributions $q(v_i) \to p(v_i) \to \tilde p(v_i) \to \cdots$.
Schematically the NN flow for each node, say for $q=2$, looks like
\be
v_j^\prt{0}\prt{=\s_j}\rightarrow h_{\a}^\prt{1}\rightarrow v_{j}^\prt{1}\rightarrow h_{\alpha}^\prt{2}\rightarrow 
...\rightarrow h_{\alpha}^\prt{n}\rightarrow v_{j}^\prt{n}~.
\ee
We start the flow by mapping the spin state $\s_i$ to the visible nodes  $v_j^\prt{0}\prt{=\s_j}$. Then we reconstruct the configurations driving them to the hidden layer $h_{\a}^\prt{1}$ and pulling them back to visible layer $v_{j}^\prt{1}$ producing a new spin microstate for the $q$-state model. We repeat the process in an iterative way several times. This consists of a generated flow by the NN, which we call it NN flow, defined in the same way for all the NN we consider as training machines.

\subsection{Autoencoder for the NN flow}

\begin{figure}
\includegraphics[angle=0,width=0.4\textwidth]{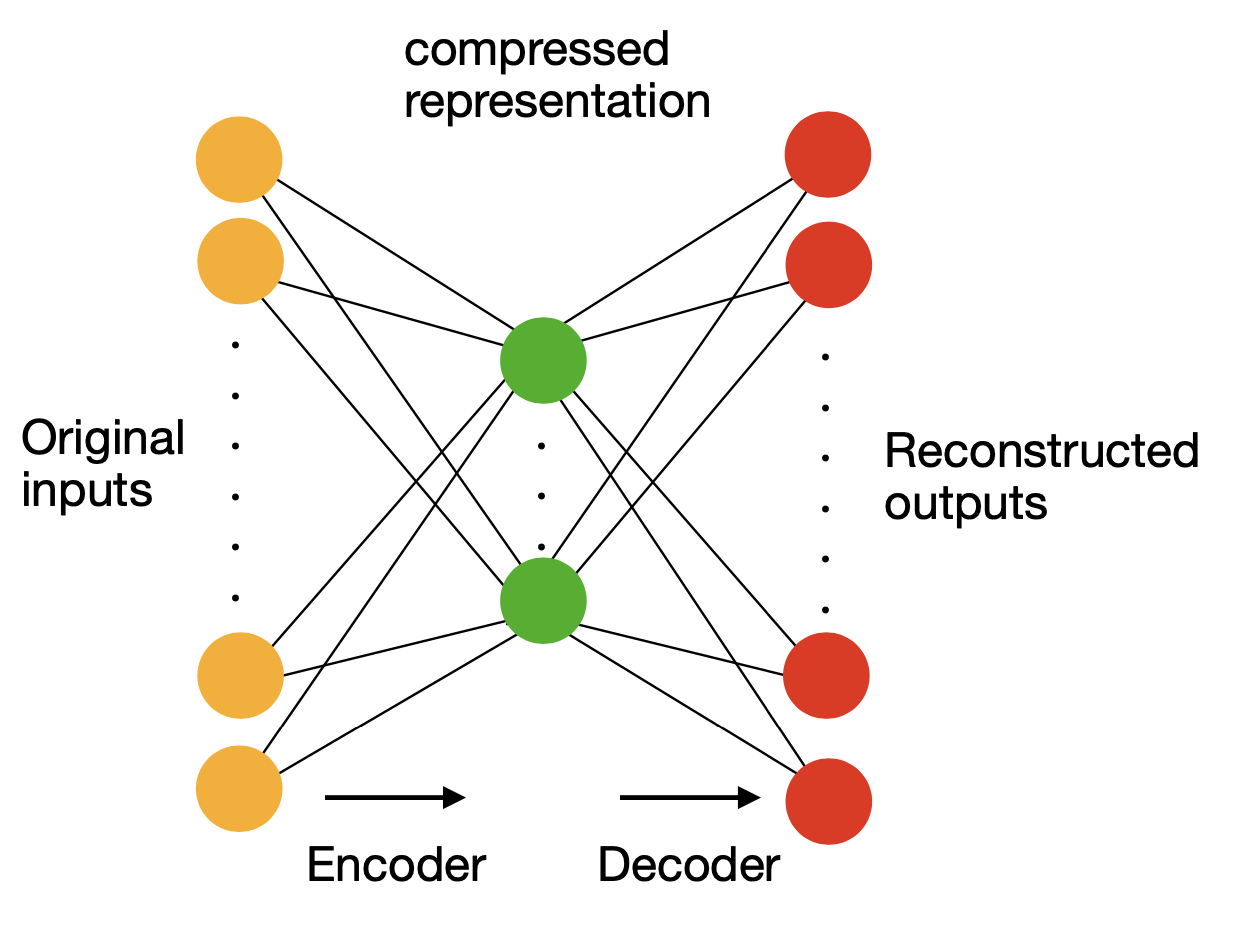}
\caption{\small Schematic structure of autoencoder (AE). the loss function for training AE is the reconstruction loss by comparing the inputs and the reconstructed outputs.}
\label{fig::ae}
\end{figure}

By its structure, the autoencoder (AE) is a natural alternative to RBM for our purpose of constructing the NN flow. The AE is a type of NN for unsupervised learning, and consists of two main parts: (1) an encoder to compress the input to into a latent vector, which is a compressed representation of the input vector. This can be seen as the reduction of the dimensions of the original input space, and is generally adopted for image compression, feature extraction and other similar tasks; (2) a decoder to try to reconstruct the input from the latent vector, which is the output of encoder. A schematic structure of the AE is shown in Fig. ~\ref{fig::ae}: an input layer to the encoder, a hidden layer to represent the compressed latent vector, and the output layer to the decoder. To optimize the AE, one should try to minimize the reconstruction loss, i.e., the difference between the input and output. Here we will use the cross entropy as the loss function, which is given by  $-\sum_{i=1}^n (x_i \log  \hat{x}_i - (1-x_i) \log(1- \hat{x}_i)$,  where $\{x_i\}$ denotes the input vector, and  $\{\hat{x}_i\}$ the output vector.

\subsection{Variational autoencoder for the NN flow}

\begin{figure}[t!]
\includegraphics[angle=0,width=0.5\textwidth]{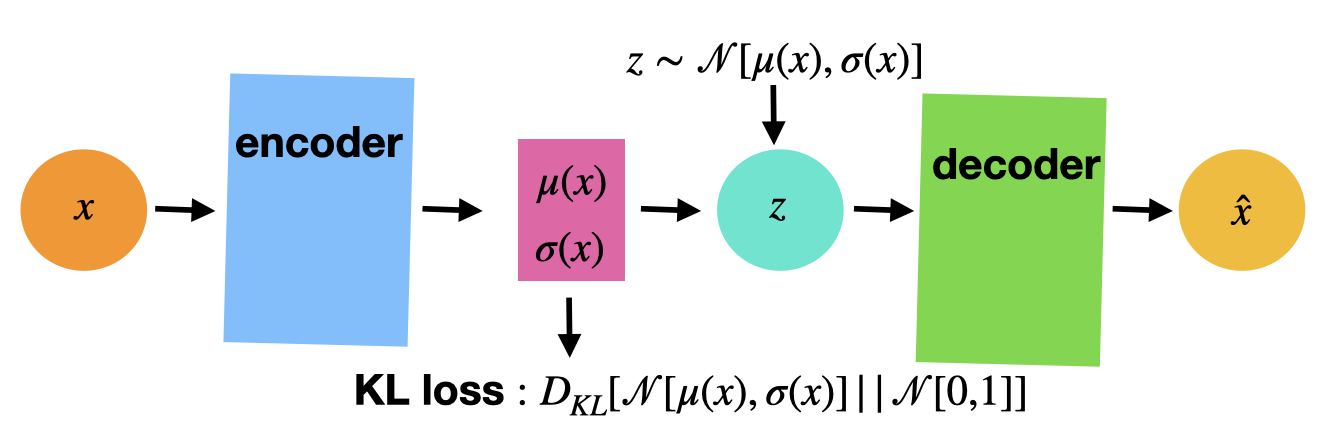}
\caption{\small The schematic structure of a variational autoencoder (VAE). The output of the encoder as the latent layer is a vector of the means and variances of Gaussian distributions, from which one can sample out an input vector $z$ to the decoder. The total loss function contains two parts: (i) the KL divergence by comparing the Gaussian of the hidden layer to a unit normal distribution; (ii) the reconstruction loss as in the case of AE.}
\label{fig:VAE}
\end{figure}

The variational autoencoder (VAE) shows almost the same machine structure as AE, namely, a VAE also contains an encoder, a latent vector and an decoder. The only difference is now that the latent vector is a random vector obeying Gaussian-like distribution. Therefore, in VAE the output of the encoder is a set of means and variances for the Gaussian distributions, from which then the latent vector is sampled and obtained. Naively, one expects the latent space distribution is a unit normal so that the deviation can be characterized by their KL divergence. The sampled latent vector as the output of the encoder is then fed into the decoder to generate the output, which can then be compared with the input to obtain the reconstruction loss.  Minimizing together the KL loss and the reconstruction loss, we can achieve the unsupervised learning. A typical schematic structure of VAE is shown in Fig. \ref{fig:VAE}. One possible advantage of VAE over AE is the flexibility to choose the latent space representation so that the map between latent vector and input vector is no longer deterministic. This turns out to be crucial for the VAE to acquire the extrapolating capability, and can be trained to be a generative model. On the other hand, the AE or VAE unit in our framework is used as the elementary step in NN flow, similar to the coarse graining step in the usual RG flow.
Thus, it is not so clear if the above difference is relevant or not.
However, it turns out that the accuracy in the machine training for the NN flow we examine, is not so relevant so that both AE and VAE usually can reach almost the same theoretically expected behaviors.

Besides, there are two key differences between the RBM machine and (V)AE machine. First, the former typically is a binary machine but the latter can be more flexible. Therefore, when applied to the higher $q$ model, we need to transform the $q$-state site into more than one binary-state sites when adopting the RBM machine.  Thus, there are quite a redundancy for representing a state in Potts model in RBM machine when $q$ is not a power of two. This will introduce more complication than dealing with (V)AE machine. Second, for simplicity one usually adopts the contrast divergence method \cite{Hinton:2002} to optimize the RBM, which is simpler in algorithm but less accurate and flexible. Otherwise, the overall procedure for training and running of the NN flow is the same as summarized in Fig. \ref{fig:Summary_of_NN_flow}.

\subsection{Neural Network Thermometer} \la{section:NNT}

\begin{figure}[t!]
\includegraphics[angle=0,width=0.4\textwidth]{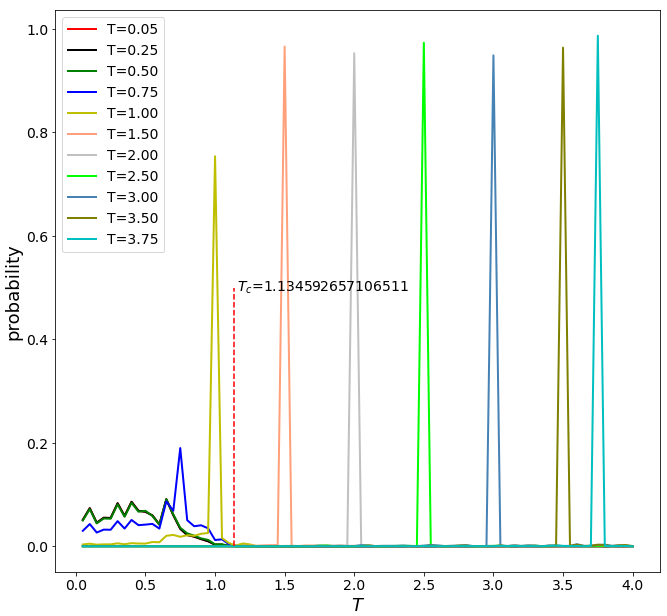}
\caption{\small The tomography of measuring the temperatures of MC training set by a typical NN thermometer . In this case, the NN thermometer is trained with MC simulation data of 2-state Potts model on a $10$ by $10$ square lattice. The dashed line indicates the critical temperature, of which there is no corresponding set.}
\label{fig:NN thermo}
\end{figure}

The generated NN flow will produce new spin configurations with a probability distribution. Our first task is to use a method to identify the temperature $T$ of the produced configuration. There are several ways to do this, a direct one is by measuring observables on the spin configuration and identifying the temperature that they correspond. Another one is to use a neural network (NN) \cite{Iso:2018yqu,giataganas1} to perform a supervised learning on the MC configurations. In this work, we have used the standard tensorflow NN to perform the supervised training for the thermometer with the same MC simulation data for training the NN flow. The test of the thermometer can yield high accuracy for most of the spin configurations except the low temperature ones, since these states tend to be nearly monochromatic.
Despite that, the regime of good accuracy usually covers the critical point so that it is suited for our purpose of tracing the NN flow to the critical point.

A typical result of the NN thermometer is shown in Fig. \ref{fig:NN thermo}. This result is obtained for the MC simulations of 2-state Potts model on a $10$ by $10$ square lattice. The inaccuracy for the low temperature regime will get improved by enlarging the lattice size. In this work we mostly consider the $20$ by $20$ square lattice, of which the NN thermometers have the better accuracy than this one. Moreover, in most of the cases considered below the thermometer has not been trained at the exact value of the appeared phase transition and it is not biased in any way.

\section{Results}\label{sec:results}

\subsection{NN flows of the 2-state models}

\begin{figure}[t!]
\includegraphics[angle=0,width=0.45\textwidth]{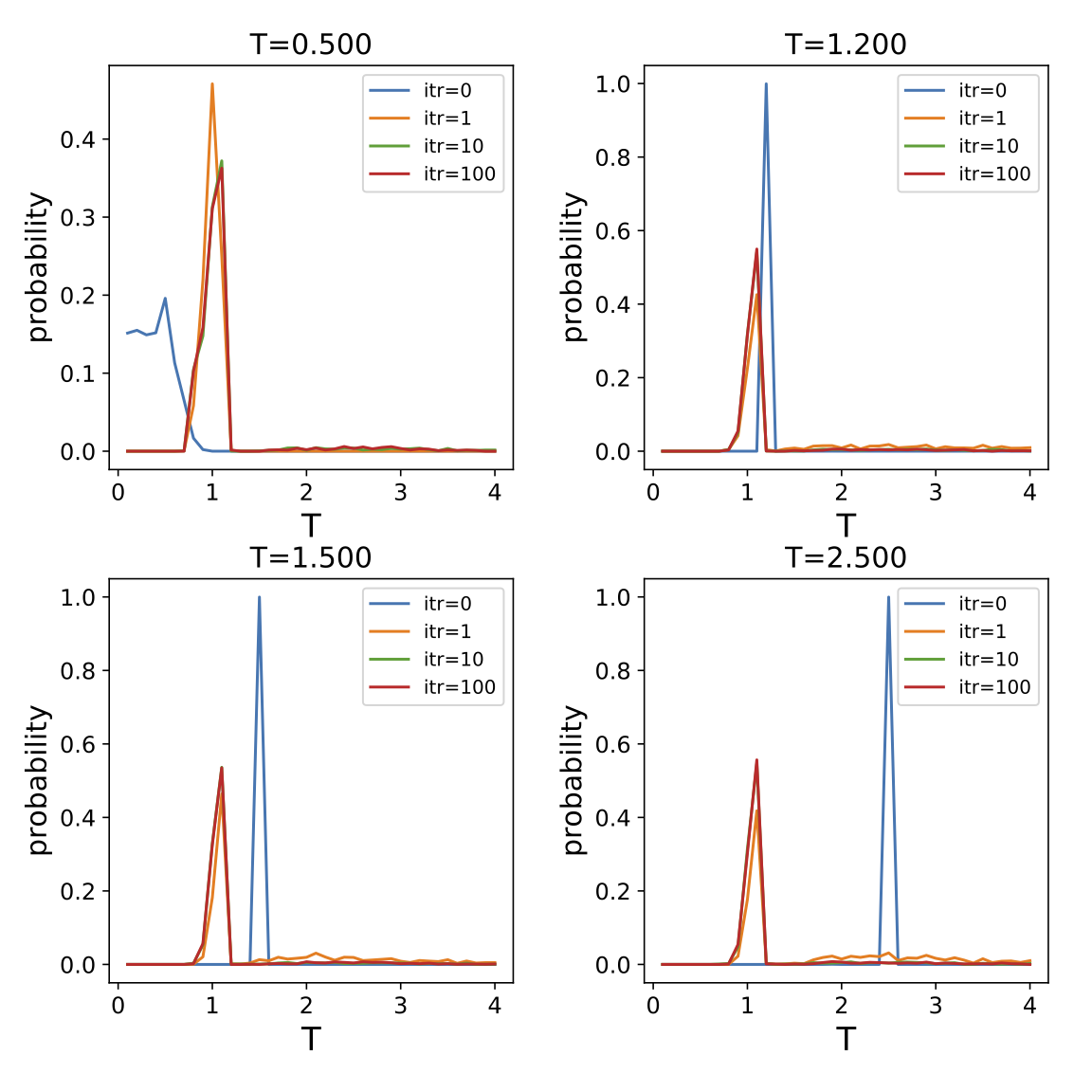}
\caption{\small RBM flow of the q=2 Potts model on a 20 by 20 square lattice. The low temperature initial configuration is not sharp, i.e., the two top sub-graphs, this is due to the inaccuracy of the NN thermometer in the low temperature regime as expected. As shown, all the indicated MC configurations flow into the same final configuration around the critical temperature $T_c\simeq1.1$ after about 10 RBM steps.
}
\label{RBM_Ising_L20}
\end{figure}

\begin{figure}[t!]
\includegraphics[angle=0,width=0.45\textwidth]{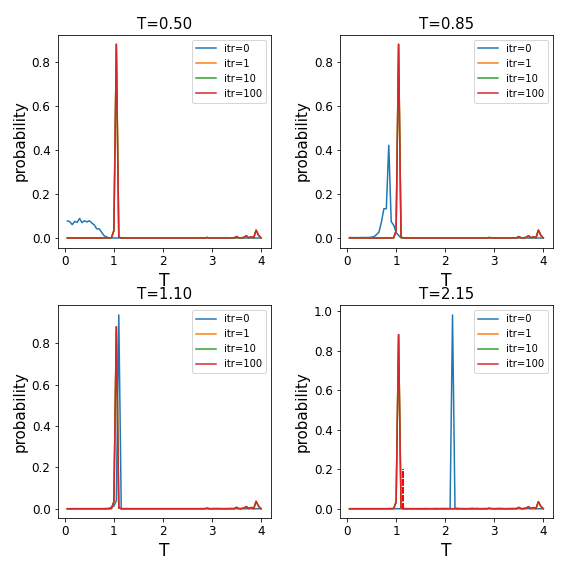}
\caption{\small VAE flow of the q=2 Potts model on a 20 by 20 square lattice.
The final configuration is  sharper around the critical temperature.
\label{VAE_Ising_L20}}
\end{figure}

We initiate our study with the $q=2$ Potts and clock models, both of which are equivalent to the Ising model, confirming that our methods converge to the Ising critical point in agreement with \cite{Iso:2018yqu,giataganas1,Koch:2019fxy}. The specific details of the MC simulation have been discussed in section \ref{sec:MC simulations}, and the corresponding phase diagrams are shown in Fig. \ref{phase-q2-clock} for the clock models as the examples.

The procedure of training the unit of NN flow and the NN thermometer, and the running of the NN flow have been sketched in Fig. \ref{fig:Summary_of_NN_flow} and described in details in the last section. The important result is that the NN flow will move toward the critical point of the system, here for Potts model the critical temperature is about $T_c=1.135$. Moreover, once it reaches the critical temperature it remains in the regime, indicating that this as a stable configuration/state of the NN flow. The RBM flow of this model is shown in Fig. \ref{RBM_Ising_L20}, and the VAE flow is shown in Fig. \ref{VAE_Ising_L20}.   The critical fixed point of the 2-state Potts model is unstable and the decimation of the degrees of freedom produces spin configurations moving away from the critical point towards the two other fixed points at $T=0$ and $T=\infty$. Therefore, the NN flow resembles an inverse of the RG flow.

From Fig. \ref{RBM_Ising_L20} and Fig. \ref{VAE_Ising_L20} we see that the flowed probability distribution settle down to its fixed point one quite quickly, i.e., usually no more than few steps. Even we just show four initial configurations, we in fact have done about 10 and all of them all flow to the same final configuration in the similar manner. In both cases we see that even the NN thermometer is not accurate in the low temperature regime as expected, the low temperature initial configuration still flow to the critical point as the high temperature ones. In some sense, the NN flow can identify the critical point very precisely even without any particular input of the critical phase transition.  In the thermodynamic limit the critical point is characterized by divergent correlation length. This feature is however weaken by the finite size effect so that the phase diagrams for some physical observables may look degenerate to a crossover, which still have some sharp but not divergent behavior.

By looking at the probability distributions of the different NN flows, we see that the VAE has sharper probability distributions than the RBM one, nevertheless they both flow to identify the same critical point.
This could be due to the ways of updating the machine. For RBM we just use the contrast divergence method, but not the standard back propagation method as we have adopted for VAE or AE from the standard tensorflow packages. The latter updating method could be more efficiency than the former one.

\begin{figure}[t!]
\includegraphics[angle=0,width=0.45\textwidth]{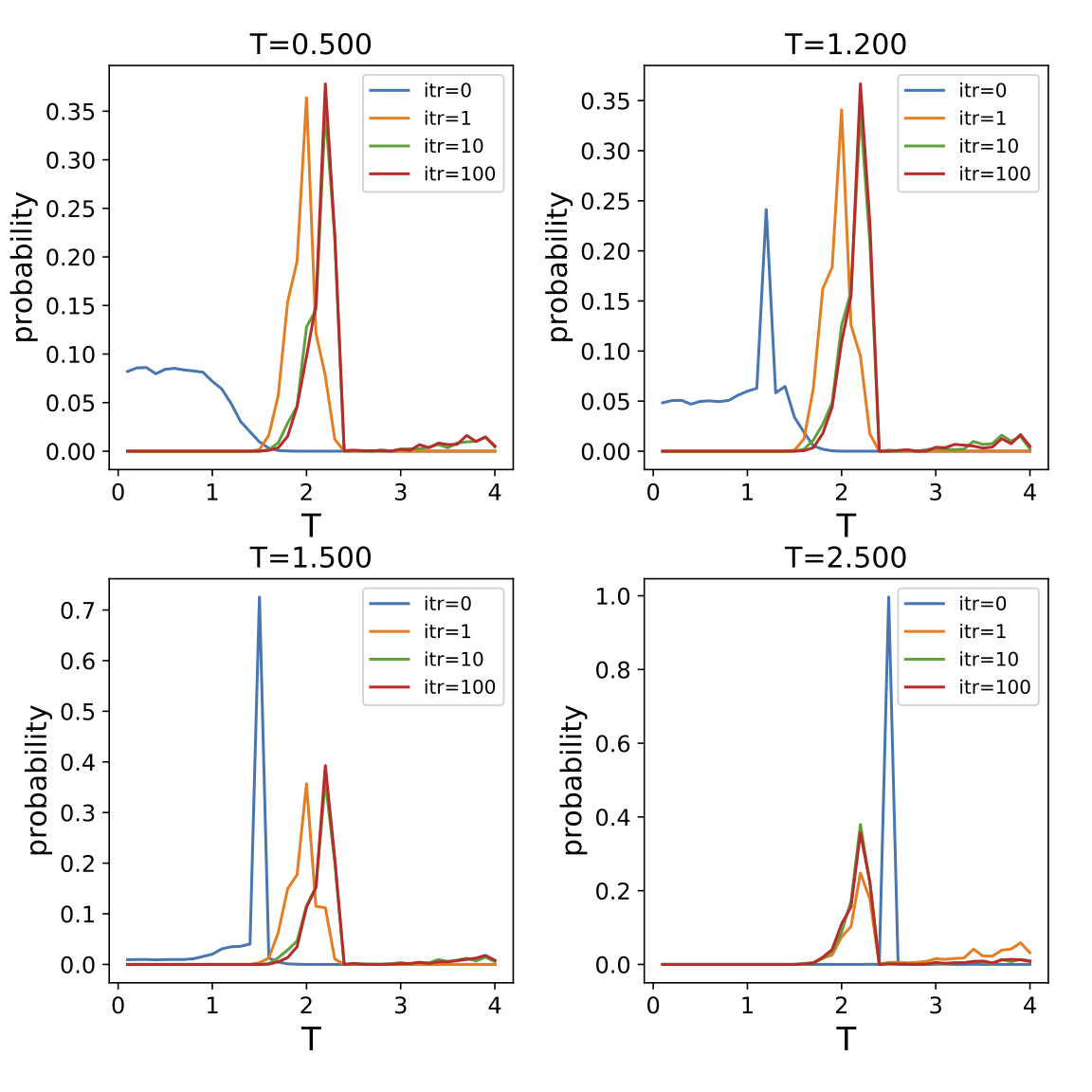}
\caption{\small RBM flow of the q=2 clock model on a 20 by 20 square lattice. As shown, after few steps of flow all the initial configurations flow to the same final one with the peak near its critical temperature $T_c\simeq2.27$.
\label{RBM_clock2_L20}}
\end{figure}

\begin{figure}[t!]
\includegraphics[angle=0,width=0.45\textwidth]{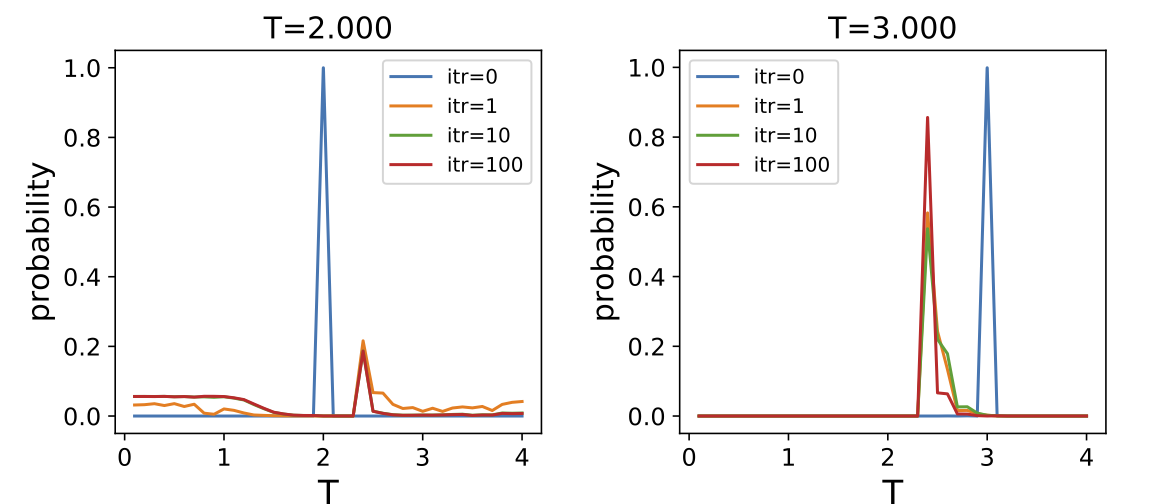}
\caption{\small AE flow of the q=2 clock model on a 20 by 20 square lattice. For simplicity, we only show the NN flow for two initial spin configurations, one below and one above $T_c$. However, all the initial configurations can be shown to flow to the same final configuration with the peak near $T_c$.
\label{q2-clock-AE}}
\end{figure}

\begin{figure}[t!]
\includegraphics[angle=0,width=0.48\textwidth]{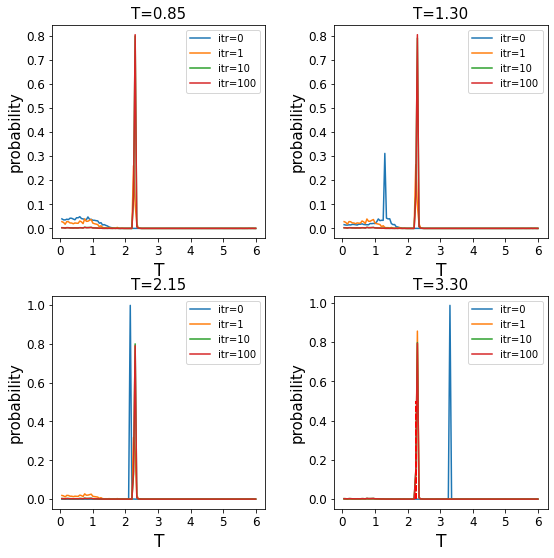}
\caption{\small VAE flow of the q=2 clock model on a 20 by 20 square lattice.  The result is the same with the RBM and AE flow, but its fixed-point configuration has a sharper peak around $T_c$.
\label{vae-rg-q2-clock}}
\end{figure}

Let us also look at the 2-state clock model. Although it is trivially equivalent to the 2-state Potts model, it is interesting to demonstrate if NN flow can work similarly by using the different training set of MC simulation spin configurations. The answer is positive, and the NN flow results for RBM, AE and VAE are shown in Fig. \ref{RBM_clock2_L20}, \ref{q2-clock-AE} and \ref{vae-rg-q2-clock}, respectively. Note that the critical temperature is twice of the Potts' one, i.e., $T_c=2.27$.
Again, we observe that the VAE has sharper probability distributions than RBM and AE, but all NNs give the same prediction for the identification of  the stable fixed point of the critical phase transition.
The sharper probability distribution may not be surprising as the VAE is a generative model as discussed before, thus it has a better extrapolating power than the deterministic NN such as RBM and AE.

One more difference from the usual machine learning related to the above discussion is that the accuracy of the training and test for the unit of the NN flow usually is quite low,
see Fig. \ref{accuracy issue} in the Appendix. However, the inaccurately trained VAE and AE flow can identify the critical point unambiguously. It seems that the pile-up flow can bypass the training accuracy to yield some generative feature.

As usual for the machine learning, the results of the NN flow still depend on the tuning of the hyperparameters. We find that the NN thermometer can be optimized for three-layer structure with about $1000$ hidden layers and run for about $400$ epochs. For the training of the NN-flow unit to be optimized for the good NN-flow results, the size of the hidden layers is about $150$ for the RBM, and $64$ for the (V)AE by running about 50 epochs. The tuning numeric of hyperparameters will vary slightly for different q-state models.

\subsection{NN flows of the 3-state models}

\begin{figure}[t!]
\includegraphics[angle=0,width=0.5\textwidth]{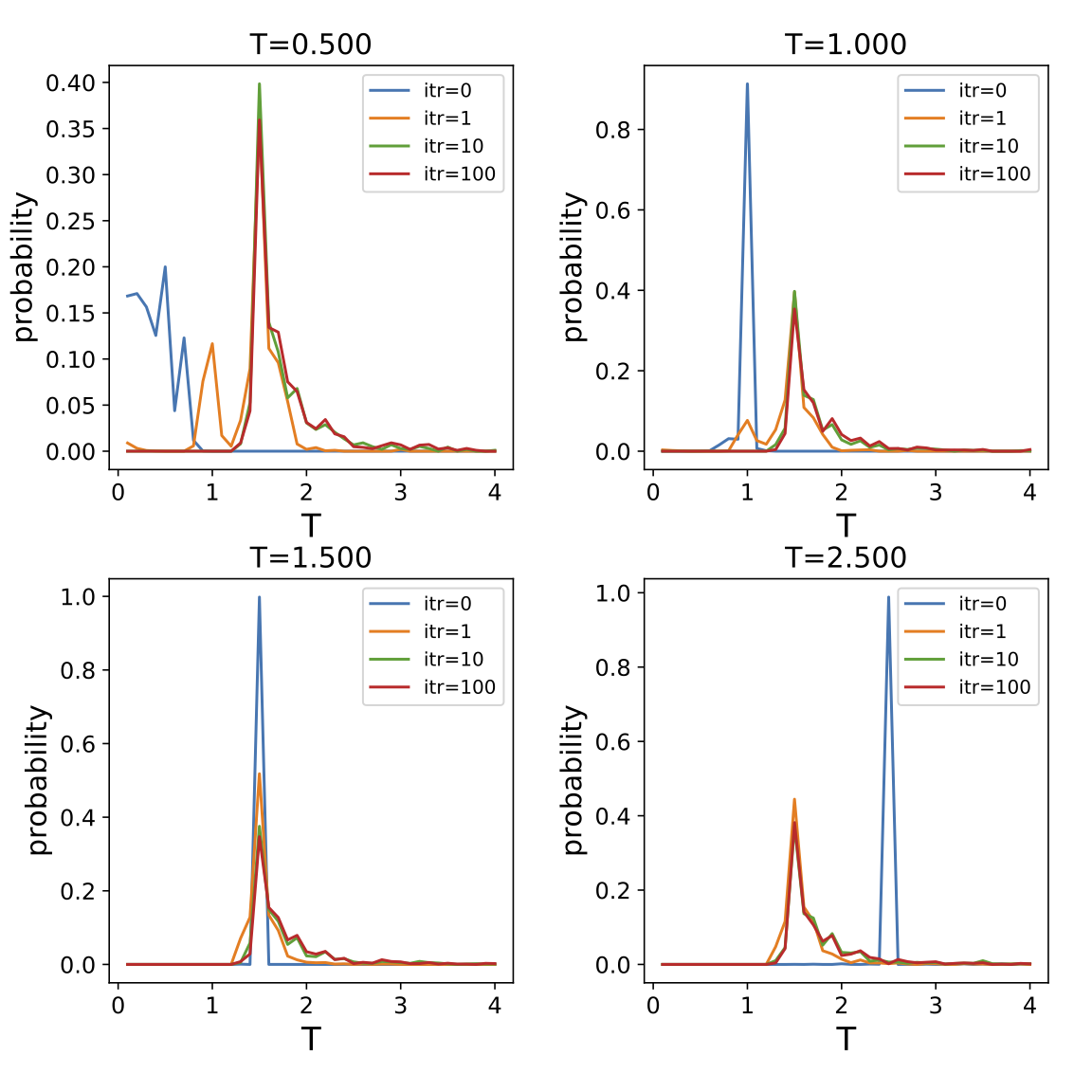}
\caption{\small RBM flow of the q=3 clock model on a 20 by 20 square lattice.  Regardless the initial spin configurations, the stable  configuration of the NN flow is peaked around the MC's critical temperature around $T\simeq 1.6$, which is slight different from the theoretical critical temperature $T_c=1.49$.
\label{RBM-rg-q3-clock}}
\end{figure}

\begin{figure}[t!]
\includegraphics[angle=0,width=0.45\textwidth]{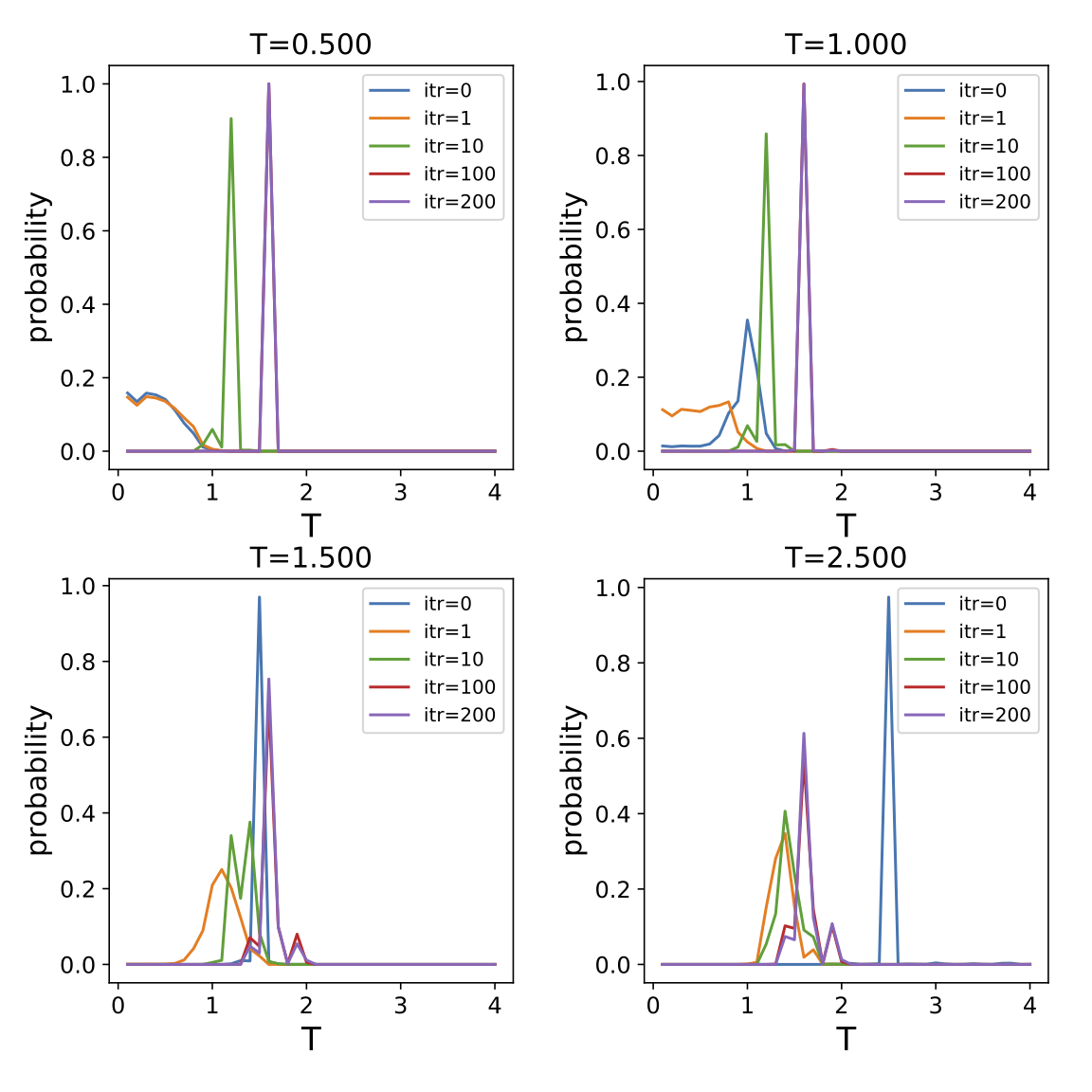}
\caption{\small AE flow of the q=3 clock model on a 20 by 20 square lattice.
The stable configuration is peaked around the MC's critical temperature at $T\simeq1.6$. }
\label{ae-rg-q3-clock-L20}
\end{figure}

\begin{figure}[t!]
\includegraphics[angle=0,width=0.45\textwidth]{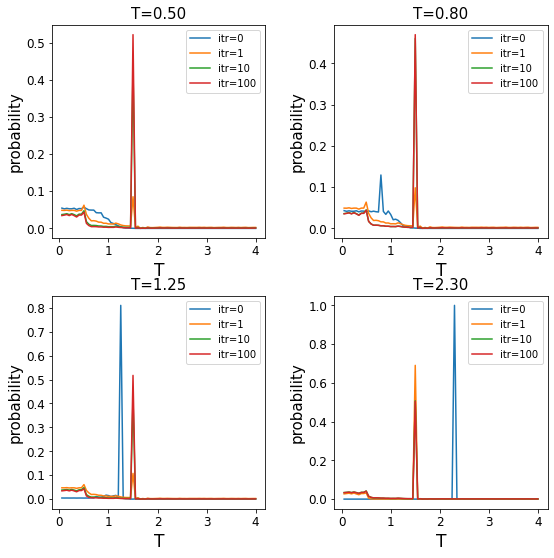}
\caption{\small VAE flow of the q=3 clock model on a 10 by 10 square lattice.
It still flows to the MC's critical point but with a sharper probability distribution than the AE and RBM ones.
\label{vae-rg-q3-clock-L10}}
\end{figure}

To examine if the aforementioned generative feature of NN flow is generic, we now move to the 3-state models. We start with the 3-state clock model, which still has a second order phase transition. The main difference compared to the binary Ising model is that we cannot map a q-state node to a binary node of RBM's visible layer, but map it to $n_q$ binary nodes where $2^{n_q}\ge q> 2^{n_q-1}$. For example, we can map the three states of a 3-state node to the $(0,0)$, $(0,1)$ and $(1,0)$ states of the two binary RBM nodes so that the $(1,1)$ state is redundant. The 3-state physical space is just the sub-space of the two binary neurons' code space on RBM's visible layer. Despite that, we do not restrict the output state to the above subspace, thus the output state can be out of the range of the 3-state physical space.
The NN thermometer adopted for the RBM flow is also implemented in the same way. On the other hand, when considering the (V)AE flow, it is not restricted to the binary NN, and the mapping between q-state physical space and the code space of the neurons is one-to-one. Thus, the implementation for the (V)AE flow is more straightforward than the RBM.

As discussed before, the NN flow should be optimized by tuning the hyperparameters, and the tuning numeric is only slightly different from the $q=2$ cases. The resultant NN flows for the 3-state clock model are in  Figs. \ref{RBM-rg-q3-clock}, \ref{ae-rg-q3-clock-L20} and \ref{vae-rg-q3-clock-L10} for the RBM, AE and VAE cases, respectively. Again we see the similar results as in the 2-state models. The NN flow can yield the stable configuration peaked around the MC's critical temperature around $T\simeq 1.6$, due to finite-size effect which is slightly different from the theoretical critical temperature of the 3-state clock model at $T=1.49$. Thus, the NN flow is capable of capturing the critical feature of the training set, i.e., the MC samples. Again, we find that the VAE yields sharp critical distribution than the RBM and AE ones. Note that the lattice size for Fig. \ref{vae-rg-q3-clock-L10} is 10 by 10, in contrast to 20 by 20 for Fig. \ref{RBM-rg-q3-clock} and Fig. \ref{ae-rg-q3-clock-L20}. This is because VAE flow fails to get optimized with the NN parameters we have tried to capture the critical point when the lattice size is enlarged. We will come back to this issue in section \ref{issues}. 

\begin{figure}[t!]
\includegraphics[angle=0,width=0.45\textwidth]{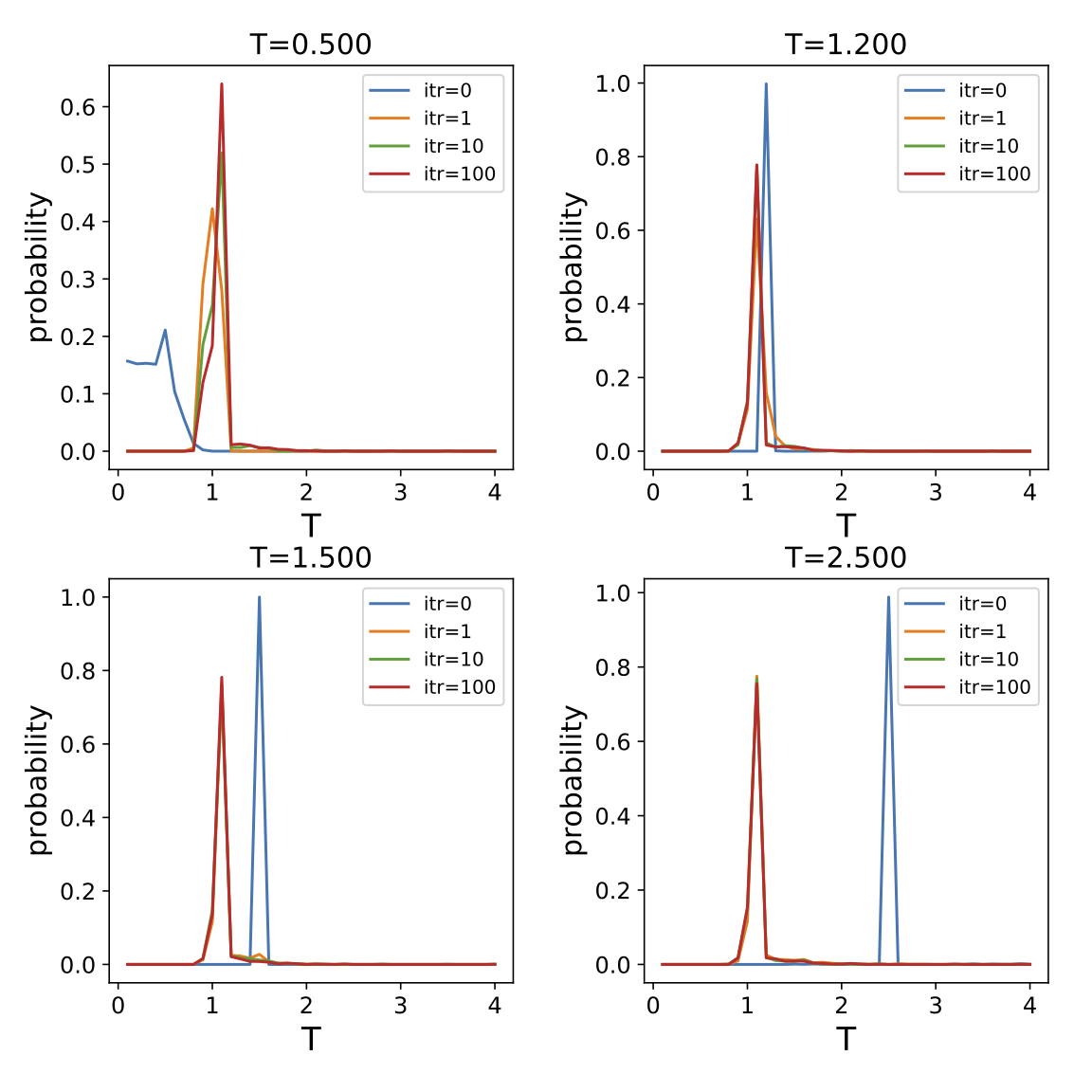}
\caption{\small RBM flow of the q=3 Potts model on a 20 by 20 square lattice. The stable state peaked around MC's critical temperature $T\simeq 1.0$ is arrived regardless the initial configurations.
\label{rbm-q3-Potts-L20}}
\end{figure}

\begin{figure}[t!]
\includegraphics[angle=0,width=0.45\textwidth]{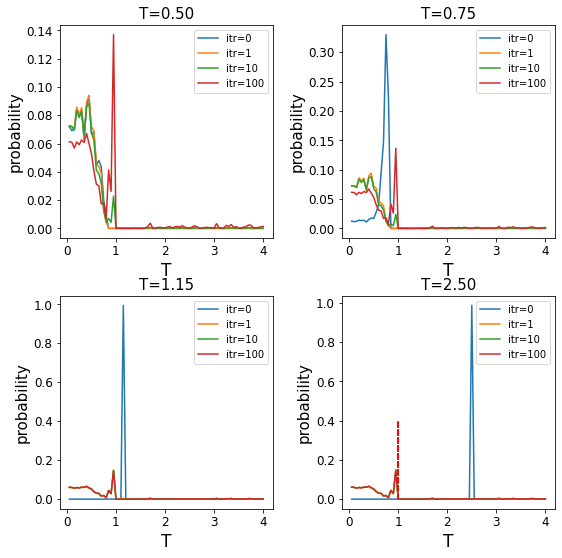}
\caption{\small VAE flow of the q=3 Potts model on a 20 by 20 square lattice. The stable state peaked around $T_c$ is arrived regardless the initial configurations.
\label{vae-rg-q3-Potts-L20}}
\end{figure}

We also study the NN flow of the 3-state Potts model, and the results are shown in Fig. \ref{rbm-q3-Potts-L20} and Fig. \ref{vae-rg-q3-Potts-L20} for the RBM and VAE cases, respectively. Again, we see that all the initial spin configurations can flow to a final configuration peaked around the MC's critical temperature at $T\simeq1.0$, which is almost the same as the theoretical one at $T_c=0.995$. In this case, the RBM flow can yield sharper critical configuration than the VAE flow.

\subsection{NN flows of the 4-state models}
We now consider the NN flows of the last $q$-state models with a second order phase transition. Note that $q=4$ is the highest $q$ for the $q$-state models to have a second order phase transition. For the higher values of $q$, the transitions of the Potts models become first order, and the ones of the clock models becomes the continuous ones, the so-called BKT phase transition \cite{Kosterlitz:1973xp,Lapilli_2006,Li_2020}.

\begin{figure}[t!]
\includegraphics[angle=0,width=0.45\textwidth]
{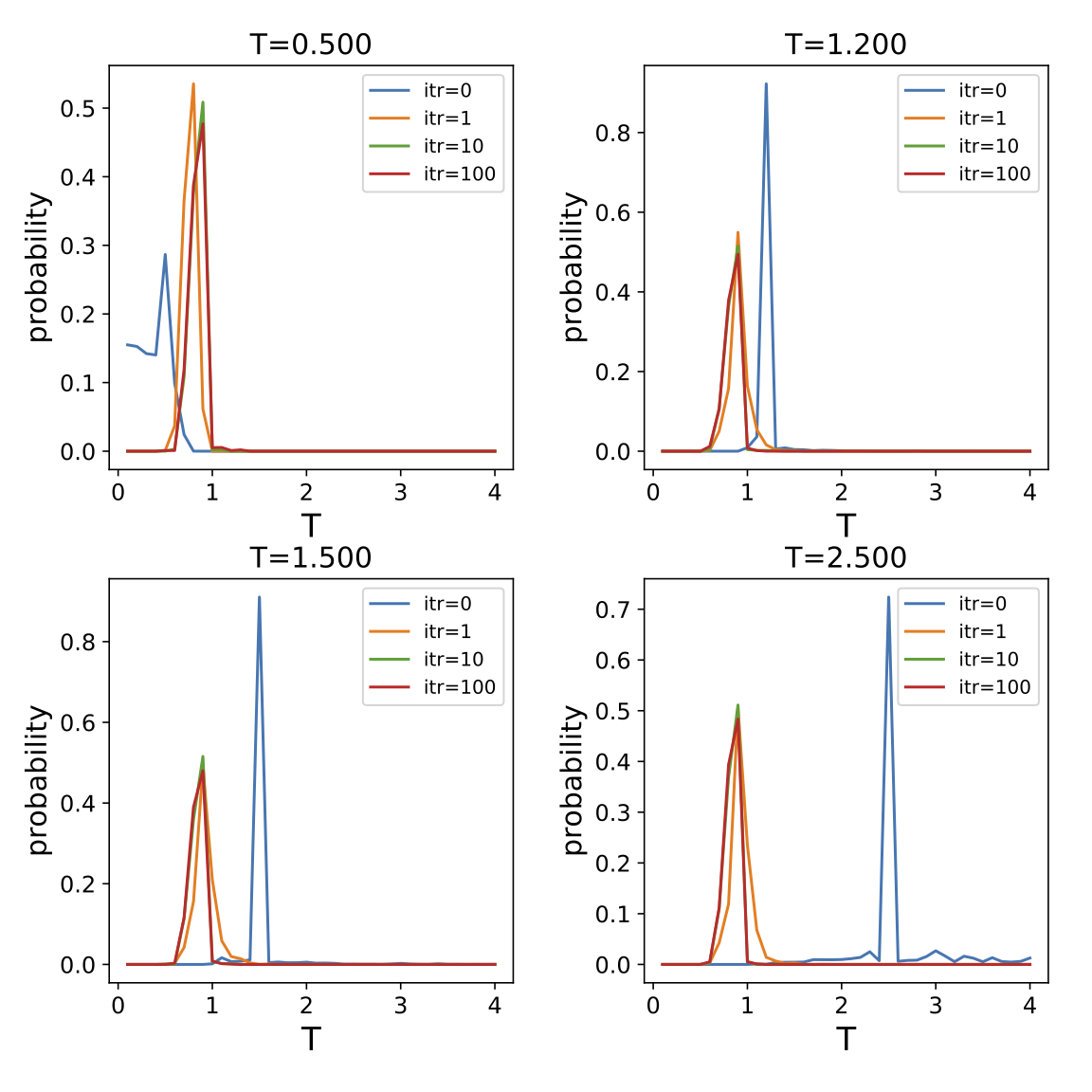}
\caption{\small RBM flow of the $q=4$ Potts model on a 20 by 20 square lattice. Despite of the increasing complexity of the physical states, it still flows to the MC's critical point.
\label{RBM-rg-q4-Potts}}
\end{figure}

\begin{figure}[t!]
\includegraphics[angle=0,width=0.45\textwidth]
{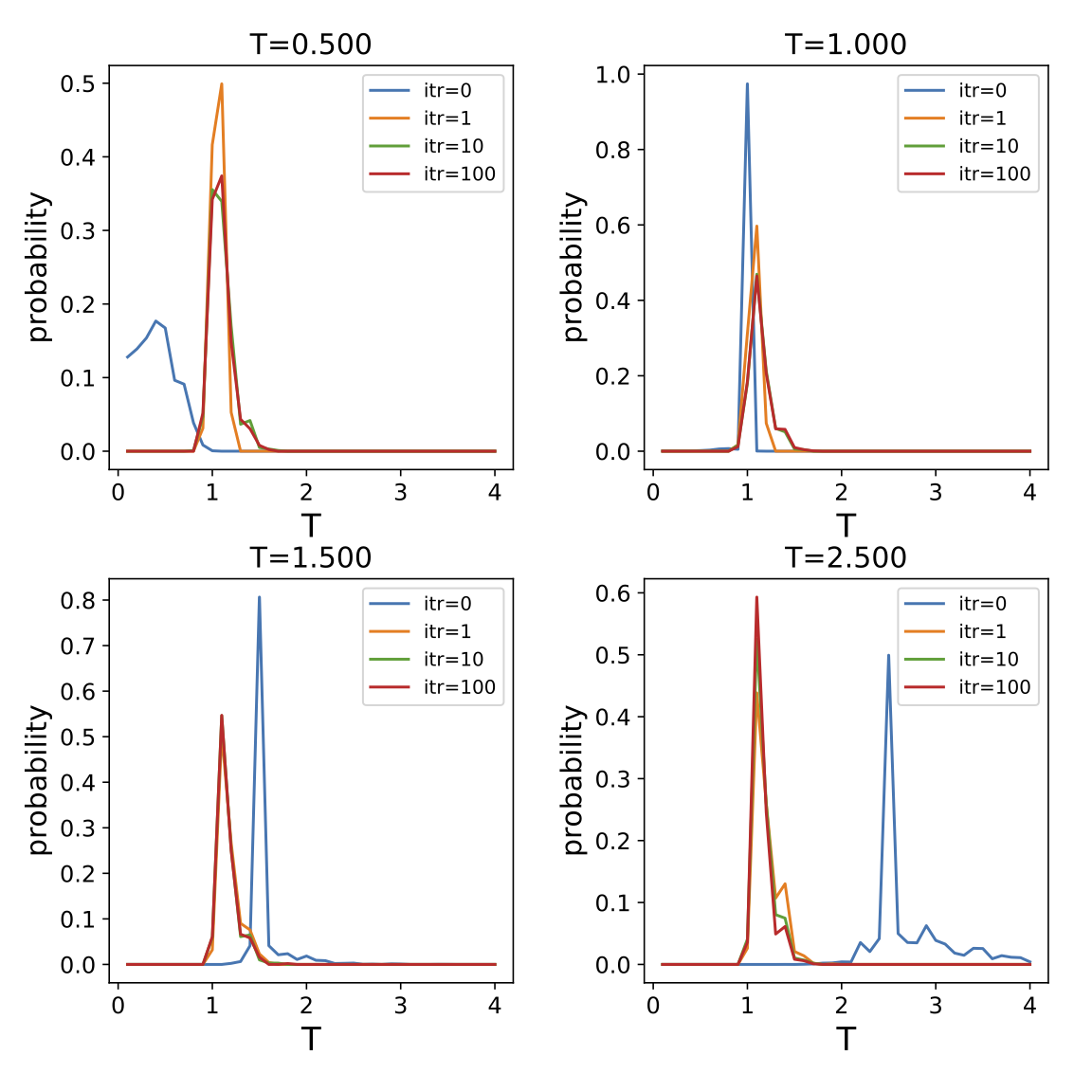}
\caption{\small RBM flow of the $q=4$ clock model on a 20 by 20 square lattice. Despite of the increasing complexity of the physical states, it flows to the MC's critical point.
\label{RBM-rg-q4-clock}}
\end{figure}

\begin{figure}[t!]
\includegraphics[angle=0,width=0.45\textwidth]{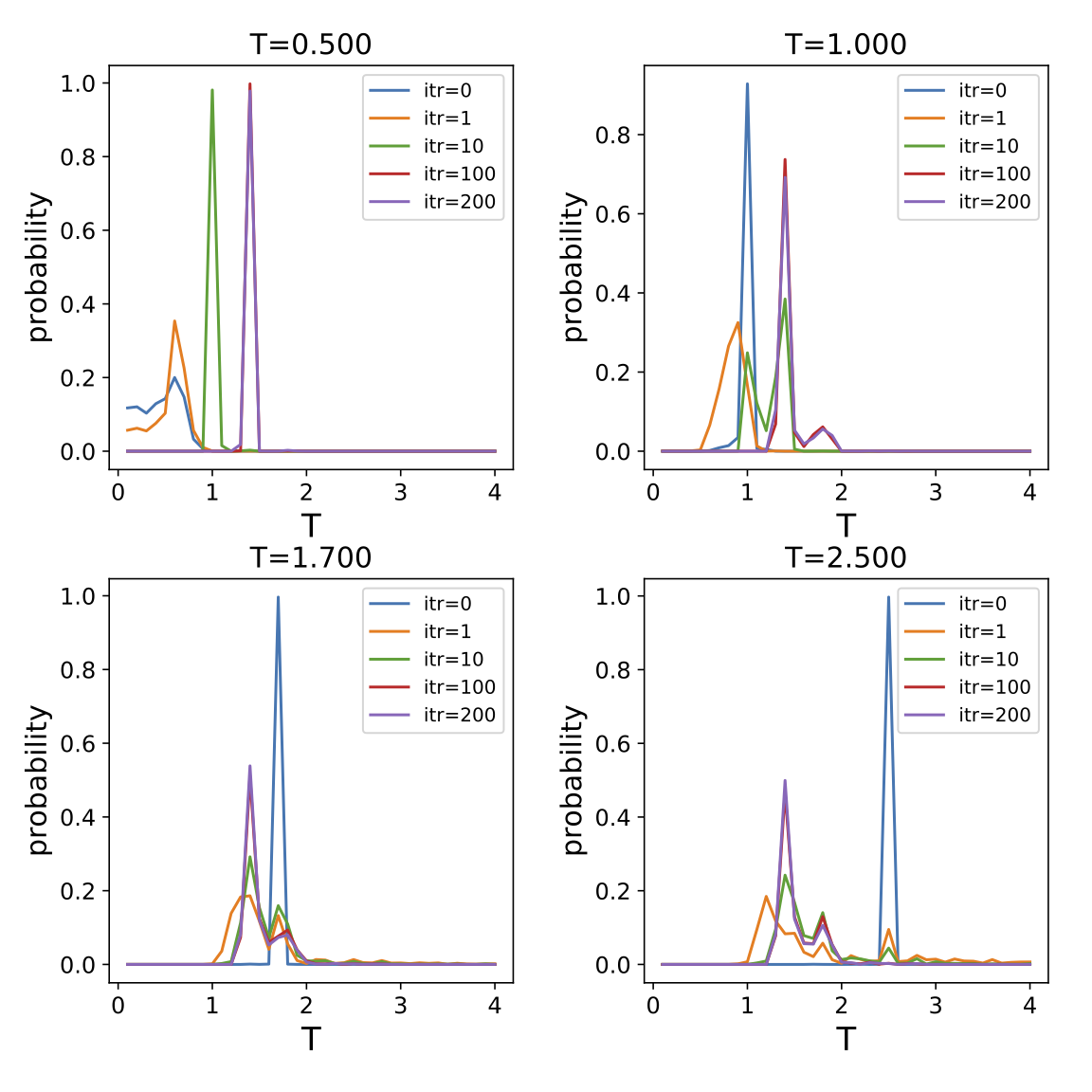}
\caption{\small AE flow of the q=4 clock model on a 20 by 20 square lattice.
The stable configuration of the AE flow is peaked around the MC's critical temperature at $T\simeq1.3$. }
\label{ae-rg-q4-clock-L20}
\end{figure}

\begin{figure}[t!]
\includegraphics[angle=0,width=0.45\textwidth]{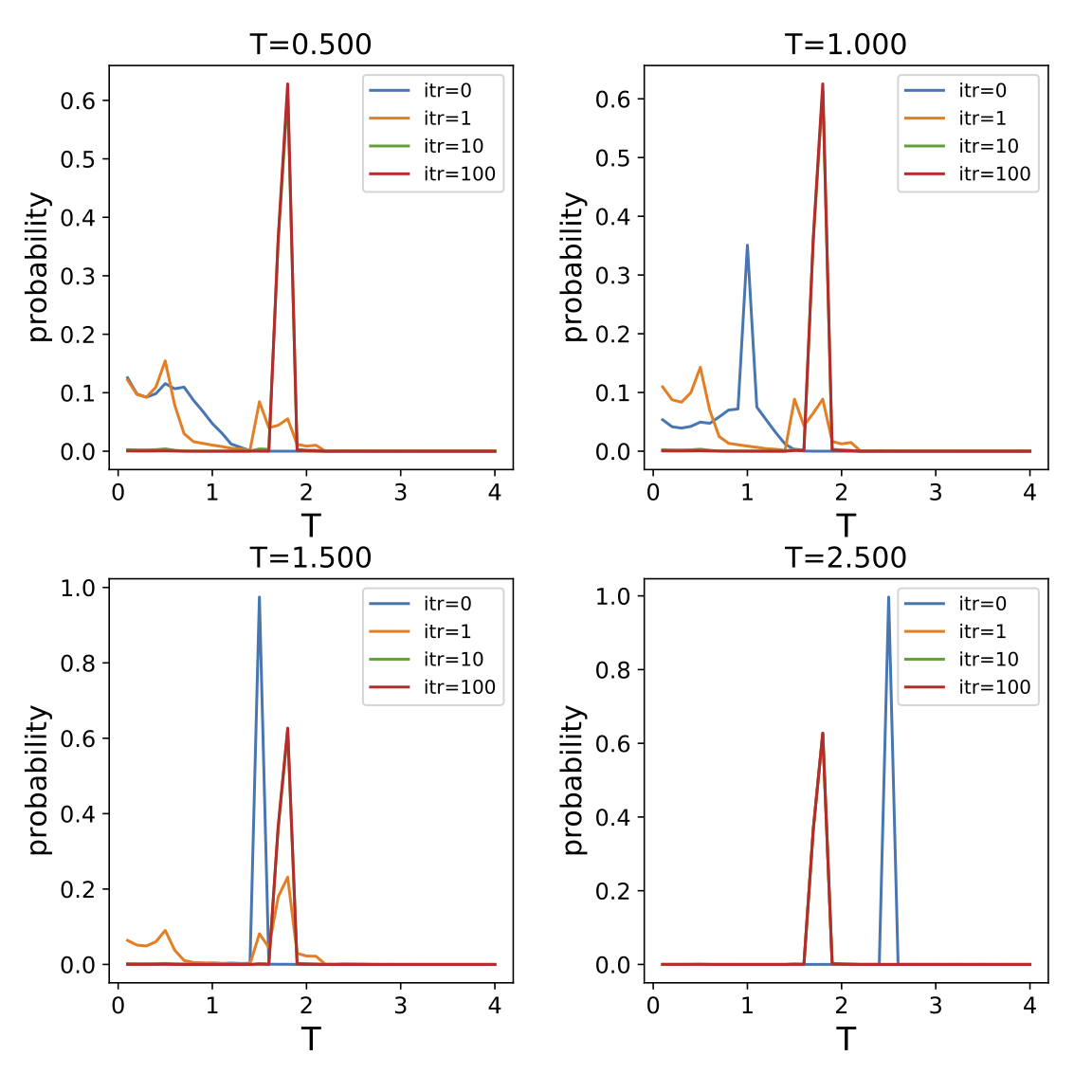}
\caption{\small VAE flow of the q=4 clock model on a 10 by 10 square lattice.
It still flows to the MC's critical point.
\label{vae-rg-q4-clock-L10}}
\end{figure}

In this case  we encode the information of the spin configuration to the machine by assigning to each node of the 4-state spin model to 2 neurons and as in the previous case we provide no other information or rules to the neural network for the mapping we have just done. Then we move on to produce the MC simulation sates, which are used for training. Once the training has finished we produce the NN flow for the RBM.  The results for the RBM flows of 4-state Potts model and clock model are respectively shown in Fig. \ref{RBM-rg-q4-Potts} and Fig. \ref{RBM-rg-q4-clock}, and the results for the AE and VAE flows of the 4-state clock model are shown respectively in Fig. \ref{ae-rg-q4-clock-L20} and Fig. \ref{vae-rg-q4-clock-L10}. As in the previous $q=2,3$ cases, all the above flow to the final configurations peaked around the MC's critical points.

\subsection{Comments on a VAE issue with complexity and on the magnetic field}\label{issues}

Before concluding our paper, we like to comment on two points regarding our NN flows. The first issue is about the complexity arising from the order of critical point. Since the clock model will turn to the continuous critical point when $q>4$, we expect that the critical point will become milder when $q$ increases. Thus, on the one hand we may need to increase the lattice size of the MC simulations to suppress the finite-size effect for better capture of the milder critical feature. On the other hand, the increasing lattice size may increase the complexity of NN-flow unit, and the difficulty in tuning the hyperparameters. This is indeed what happens when we consider the NN flow of the $q=3,4$ clock models. Although both the RBM and AE flows can capture well the critical feature for 20 by 20 lattice as shown in Fig. \ref{RBM-rg-q3-clock}, \ref{ae-rg-q3-clock-L20}, \ref{RBM-rg-q4-clock} and \ref{ae-rg-q4-clock-L20}, the VAE flow will yield either final configurations peaked around low temperature or high temperature regimes by extensive tuning of hyperparameters. On the other hand, the VAE flow for the $q=3,4$ clock models on 10 by 10 lattice can yield the expected critical configurations as shown in Fig. \ref{vae-rg-q3-clock-L10} and \ref{vae-rg-q4-clock-L10}. It seems that the generative feature of VAE does not help in dealing with the increasing complexity and possibly requires a very sensitive tuning of NN hyperparameters in order to develop a stable point at finite non-zero temperatures. 

\begin{figure}[t!]
\includegraphics[angle=0,width=0.45\textwidth]{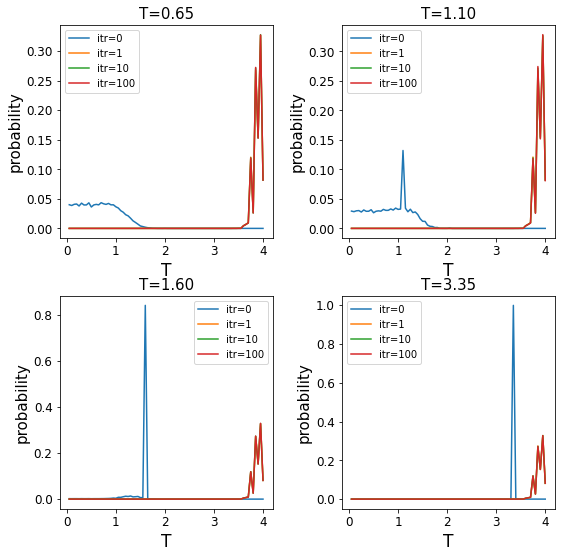}
\caption{\small VAE RG flow of the q=2 clock model on a 20 by 20 square lattice with magnetic field $h=0.05$, trained only at this value. The line style for the NN flow is the same as in Fig. \ref{vae-rg-q2-clock}. In contrast to Fig. \ref{vae-rg-q2-clock}, the MC  configurations flow to the high temperature regime, for the range of the NN hyperparameters we have attempted the training.
\label{vae-rg-q2-clock-h}}
\end{figure}

The second point is about the capture of critical behavior by NN flow after turning on the magnetic field, i.e., adding the following term to the Hamiltonian,
\be
-h \; \sum_i \cos \theta_i.
\ee
This term will pick up a preferred spin direction and destroy the $Z_2$ symmetry to lift its spontaneous symmetry breaking and to yield a crossover near the critical point of the cases without magnetic field. It is interesting to ask if the NN flow can capture the crossover or not. With extensive tuning of hyperparameters, in Fig. \ref{vae-rg-q2-clock-h} we show a VAE flow for the 2-state clock model by turning on  $h=0.05$ and training the NN only at this particular value of magnetic field, we see that the NN flow reaches a final state peaked around the high temperature regime, at least for the range of the NN hyperparameters we have used for the training.  Similar results can be  obtained for other choices of magnetic fields.

However, the training done with a fixed value of a magnetic field is an oversimplified choice. The magnetic field is a dimensionful quantity and for the purposes of the NN flow the system  should be trained in a finite range of values as in \cite{giataganas1} where the training set includes the MC configurations of different temperatures and magnetic fields so that the RBM flow can reach a final configuration peaked around the crossover temperature.  This implies that the key feature of the crossover associated with magnetic field is encoded in the variation of the magnetic field as expected.

\section{Discussion and Conclusion}\label{sec:con}

In this paper we have provided clear evidence that the NN flows develop a stable point for spin models with increased complexity. In particular we have used three different machines the RBM, the AE and VAE, to generate the NN flow in $q=2,3,4$ Potts model. Irrespectively of the machine used we have found that the flow approaches spontaneously to a stable point, which matches the critical point of the corresponding physical system. In other words the NN flows, develops a universal behavior among different physical spin models and can be used as powerful tools to identify the criticality of spin systems.

The NN flow is defined as the iterative reconstructions of data once the machine has been trained. The convergence on the process depends on the hyperparameters of the training machine, like number of the hidden nodes, but once these are optimized, there is finite neighboring regime that generate the same flow. In this sense the flow depends on the hyperparameters but is weakly sensitive on them. Moreover the fixed point of the flow is independent of the initial starting point spin configuration that generates the flow. This consists of a further evidence of the universality of the NN flow.

Our results can aid to the direction of providing a link between the RG flow and certain machine learning processes. We show that the NN flow develops spontaneously a stable point that matches with the critical point of the multistate-spin models, which suggested that the NN flows can extract the feature of critical point such as scale invariance or maximizing certain thermodynamic quantities, which are also the feature of the fixed point of RG flow. Such a connection between RBM and the RG flow other than using the NN flow approach \cite{Iso:2018yqu,giataganas1} has also been discussed \cite{mehta1,Tegmark,Koch:2019fxy,Chung:2020jyo}.  Our findings in this work call for further exploration about this connection.

\begin{acknowledgments}
D.G would like to thank A. Athenodorou, R. de Mello Koch and F. Diakonos for useful discussions and the Institute of Accelerating Systems and Applications (IASA) in Athens.  D.G is supported by the Hellenic Foundation for Research and Innovation (H.F.R.I.) and the General Secretariat for Research and Technology (GSRT), under grant agreement No 2344.
CYH is supported by Taiwan Ministry of Science and Technology through Grant
MOST 108-2112-M-029-006-MY3. FLL is supported by Taiwan Ministry of Science and Technology through Grant No.~109-2112-M-003-007-MY3, and he also thanks Han-Shiang Kuo and Chung-Hao Liao for discussions on VAE and the support from NCTS.
\end{acknowledgments}

\appendix*
\section{Training accuracy and typical layer-views of AE and VAE}\label{Appendix}

\begin{figure}[H]
\centering
{
\includegraphics[height=0.20\textwidth]{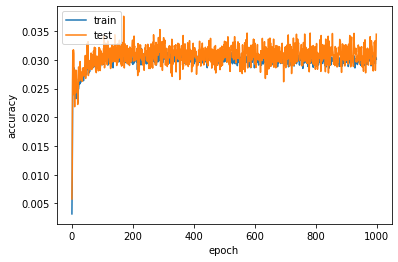}
\quad~~~
\includegraphics[height=0.20\textwidth]{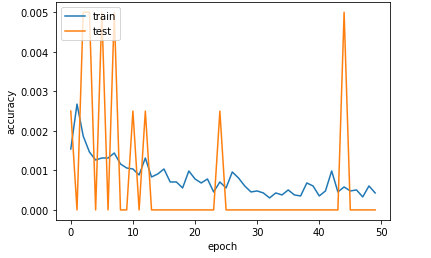}
}
\caption{\small Typical training and test accuracy of AE (Top) and VAE (Bottom) for the corresponding NN flows of Fig. \ref{q2-clock-AE} and \ref{vae-rg-q2-clock}, respectively.
}
\label{accuracy issue}
\end{figure}

\begin{figure}[H]
\includegraphics[height=0.09\textwidth]{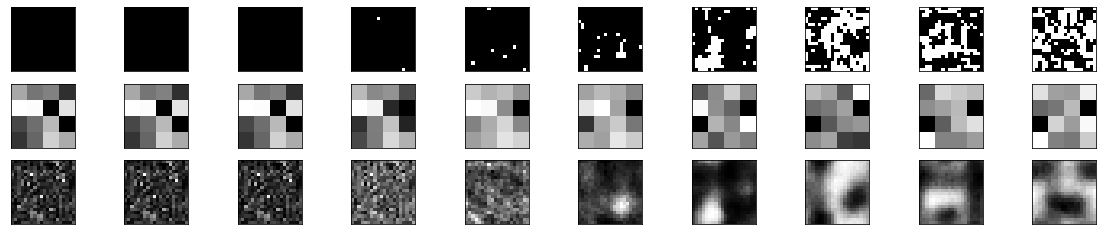}
\caption{\small A typical AE site view of the input (top), latent (middle) and output (bottom) from the corresponding AE flow of Fig. \ref{q2-clock-AE}.
\label{ae-site-q3-clock}}
\end{figure}

\begin{figure}[H]
\includegraphics[height=0.16\textwidth]{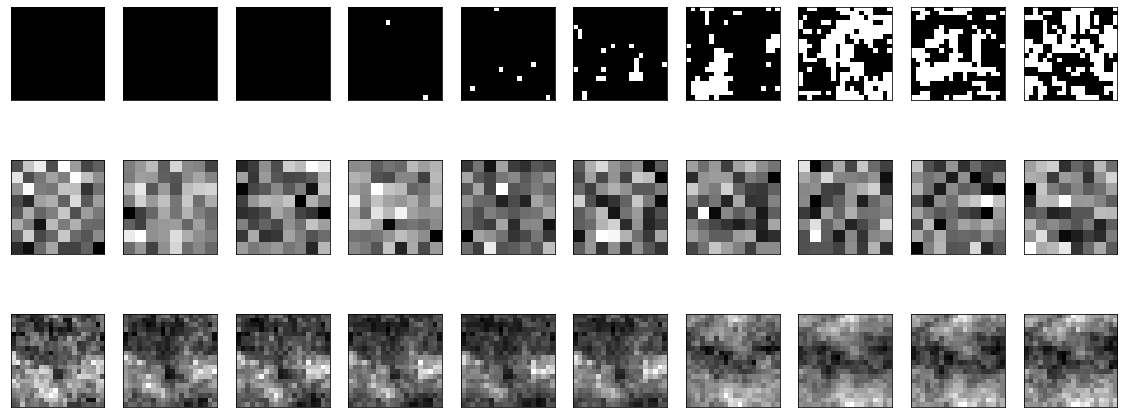}
\caption{\small  A typical VAE site view of the input (top), latent (middle) and output (bottom) from the VAE flow shown in Fig. \ref{vae-rg-q2-clock}. Compared to the AE ones in Fig. \ref{ae-site-q3-clock}, the latent ones are blurred and the output ones are more distinguishable.
\label{VAE-site-q3-clock}}
\end{figure}

In this appendix, we show a typical training accuracy for training the unit of AE and VAE flow, see Fig. \ref{accuracy issue}. The way we use the NN flow for our purposes and the convergence of the NN flows to the critical regime seems to be independent of the low accuracy. Besides, we also show some typical site-view of the layers of the AE and VAE in Fig.\ref{ae-site-q3-clock} and  \ref{VAE-site-q3-clock}, respectively. We can see the fuzzy latent layers of VAE in comparison with the deterministic ones of the AE.


~

%

\end{document}